\documentclass{arxiv}
\usepackage{helvet}
\usepackage{cite}
\usepackage{kantlipsum, lipsum}
\usepackage[pagebackref=false,breaklinks=false,%
            colorlinks=true,bookmarks=true,citecolor=ourdarkblue,%
            urlcolor=ourdarkblue,linkcolor=ourdarkblue]{hyperref}
\usepackage{multibib}

\usepackage{dm-colors}
\usepackage{amsmath}
\usepackage{pstricks, pst-node}
\usepackage{verbatim}
\usepackage{multirow}
\usepackage{array}
\usepackage{longtable}
\usepackage{pdflscape}
\usepackage{scalerel}
\usepackage{booktabs}
\usepackage{enumitem}
\usepackage{xspace}
\usepackage{bm}
\usepackage{bbm}
\usepackage{booktabs}
\usepackage{tabularx}
\usepackage{array}
\usepackage{mathtools}
\usepackage{soul}
\usepackage{epsfig}
\usepackage{graphicx}
\usepackage{amssymb}
\usepackage{colortbl}
\usepackage{csquotes}
\usepackage{setspace}
\usepackage{bbding}
\usepackage{siunitx} 
\usepackage{threeparttable}
\usepackage{tabularx,ragged2e}
\usepackage{placeins}
\usepackage{bbding}
\definecolor{darkpastelgreen}{rgb}{0.13, 0.55, 0.13}
\definecolor{darkpastelred}{rgb}{0.55, 0.13, 0.13}
\definecolor{mygray}{rgb}{1, 1, 1}
\usepackage{lmodern}
\usepackage[hang,flushmargin]{footmisc}
\usepackage{nameref}
\usepackage{varioref}
\usepackage{amssymb}
\usepackage{pifont}
\usepackage{booktabs}
\usepackage{tabularx}
\usepackage{array}
\usepackage{threeparttable}
\usepackage{rotating}
\usepackage{graphicx}

\usepackage[noabbrev,capitalize]{cleveref}
\usepackage{etoc}
\usepackage{tikz}

\usepackage{wasysym}
\usepackage{algorithm}
\usepackage{makecell} 
\usepackage{algpseudocode}
\usepackage{xcolor}
\usepackage{wasysym}  
\usepackage[most]{tcolorbox}

\newtcolorbox{casestudy}[2][]{%
  colback=gray!5,colframe=gray!50,
  fonttitle=\bfseries,
  title={Case Study: #2},
  breakable,enhanced,
  left=1em,right=1em,top=0.8em,bottom=0.8em
}

\usepackage{amsmath}
\usepackage{thmtools}
\usepackage{wasysym}
\declaretheoremstyle[
    spaceabove=6pt, spacebelow=6pt,
    headfont=\bfseries, headpunct={.}, headformat={\NAME\ \NUMBER},
    bodyfont=\normalfont,
    postheadspace=0.5em
]{promptstyle}

\tcolorboxenvironment{prompt}{
    colback=gray!10!,
    colframe=gray!75!,
    fonttitle=\bfseries,
    title=Prompt,
    boxrule=0.5pt,
    sharp corners
}

\graphicspath{{figures/}}
\definecolor{mygray}{rgb}{0.85, 0.85, 0.85}

\definecolor{lightpink}{RGB}{255,235,238}
\definecolor{lightgreen}{RGB}{232,245,233}

\usepackage{listings}

\usepackage{svg}

\usepackage{lineno}

\definecolor{codegreen}{rgb}{0,0.6,0}
\definecolor{codegray}{rgb}{0.5,0.5,0.5}
\definecolor{codepurple}{rgb}{0.58,0,0.82}
\definecolor{backcolour}{rgb}{0.95,0.95,0.92}
\definecolor{framecolor}{rgb}{0.8,0.8,0.8}

\lstdefinestyle{prettyjson}{
    backgroundcolor=\color{backcolour},   
    commentstyle=\color{codegreen},
    keywordstyle=\color{blue}\bfseries,
    numberstyle=\tiny\color{codegray},
    stringstyle=\color{codepurple},
    basicstyle=\ttfamily\small,
    breakatwhitespace=false,         
    breaklines=true,                 
    captionpos=b,                    
    keepspaces=true,                 
    numbers=left,                    
    numbersep=8pt,                  
    showspaces=false,                
    showstringspaces=false,
    showtabs=false,                  
    tabsize=2,
    frame=single,
    frameround=tttt,
    framerule=0.5pt,
    rulecolor=\color{framecolor},
    xleftmargin=15pt,
    xrightmargin=15pt,
    aboveskip=15pt,
    belowskip=15pt,
    columns=flexible,
    escapeinside={(*@}{@*)}
}

\title{\Large{Evaluating Scaffolding-Oriented Multi-Agent Large Language Model System for Clinical Interview Training
}}

\author[1]{Luming Yang}
\author[2]{Haoxian Liu}
\author[3]{Siqing Li}
\author[4]{Rong Jia}
\author[5]{Yue Xiao}
\author[3]{Guanhua Chen}
\author[6,$\dagger$]{Li Lu}

\affil[1]{\small Department of Electrical and Computer Engineering, The Ohio State University, Columbus, OH, USA \authorcr \vspace{0.1cm}}

\affil[2]{\small Department of Computer Science and Engineering, The Hong Kong University of Science and Technology, Hong Kong SAR, China \authorcr \vspace{0.1cm}}

\affil[3]{\small Department of Statistics and Data Science, Southern University of Science and Technology, Shenzhen, China \authorcr \vspace{0.1cm}}

\affil[4]{\small School of Education, Johns Hopkins University, Baltimore, MD, USA  \authorcr \vspace{0.1cm}}

\affil[5]{\small School of Medicine, Southern University of Science and Technology, Shenzhen, China \authorcr \vspace{0.1cm}}

\affil[6]{\small School of Basic Medicine, Guangzhou Medical University, Guangzhou, China \authorcr \vspace{0.1cm}}

\affil[$\dagger$]{\small Corresponding author. luli@gzhmu.edu.cn \authorcr \vspace{0.1cm}}

\begin{document}
\doublespacing

\begin{abstract}

Clinical education must prepare medical students to conduct safe, coherent, and patient-centered interviews under conditions of uncertainty. Traditional standardized patient (SP) training is resource-intensive and difficult to scale, while case-based learning alone does not reproduce the real-time communicative demands of a consultation.
We developed a scaffolding-oriented multi-agent Large Language Model (LLM) AI Standardized Patient (AI-SP) training platform\footnotemark. The system includes a patient agent for simulated dialog, a tutor agent providing Socratic prompts for history taking, clinical reasoning, and empathic communication without disclosing diagnostic information, and a turn-level evaluator agent that monitors clinical progress without revealing summative scores.
In a randomized controlled study (N = 100 medical students), participants were assigned to either a multi-agent (MA) scaffolding condition or a structured non-LLM control (CT) condition based on progressive information disclosure. All students completed two learning sessions under their assigned condition followed by an examination conducted in a patient-only environment. Performance was assessed using a standardized Objective Structured Clinical Examination (OSCE)-aligned rubric.
While no significant difference was observed in final diagnostic accuracy between groups, the multi-agent AI standardized patient system improved final examination scores compared with the structured control condition; the most substantial and consistent improvements were observed in communication, observable empathic expression, and specific history-taking behaviors. These findings suggest that specialized LLM agents can enhance the process quality of simulated clinical interviews without artificially inflating diagnostic endpoints.
To support future research, we release a multi-expert annotated dataset comprising transcripts, checklist annotations, turn-level evaluations, and OSCE-aligned scoring outcomes. This resource aims to facilitate the development of pedagogically grounded AI-SP systems and advance research on AI-supported clinical interview training.

\end{abstract}

\maketitle

\section{\texorpdfstring{Introduction}{Introduction}}

\footnotetext{Code is available at: \url{https://github.com/skylynf/agent-medu}}

Medical education must prepare students to conduct clinical interviews that are safe, coherent, diagnostically useful, and patient centered. Clinical interviewing competence includes several interdependent behaviors: gathering structured information, interpreting cues under uncertainty, generating and revising diagnostic hypotheses, and communicating in ways that acknowledge patient concerns and support disclosure. Empathic communication is therefore not peripheral to clinical reasoning; it shapes rapport, patient cooperation, the completeness of elicited information, and the observable quality of the consultation \cite{mcbee_context_2018}. Because diagnostic work often takes place under time pressure and incomplete information, students need repeated opportunities to practice both the cognitive and communicative behaviors through which clinical reasoning is enacted \cite{dai2026development, braun_representation_2017}.

Case-based learning (CBL), computerized cases, and virtual patients can support clinical and diagnostic reasoning training by allowing learners to analyze cases, compare hypotheses, and receive structured feedback \cite{cook_computerized_2010, braun_representation_2017}. Simulated patient encounters serve a complementary purpose: they require learners to enact reasoning through real-time history taking, question sequencing, explanation, rapport building, and patient-centered communication. Simulation-based learning can support complex skill development when activities incorporate appropriate scaffolding \cite{chernikova_simulation-based_2020}, and technology-enhanced simulation has been used to foster clinical skills, critical thinking, learner confidence, and patient-safety competencies \cite{barry_issenberg_features_2005, stenseth_simulation-based_2025}. However, traditional standardized patient training is difficult to scale because it typically requires trained actors, faculty supervision, physical space, fixed scheduling, and substantial institutional resources \cite{nestel_key_2011, gillette_cost-effectiveness_2017}. These constraints limit opportunities for repeated practice, timely feedback, and exposure to diverse clinical cases.

Recent digital and AI-supported approaches have attempted to address these constraints. Computerized and virtual-patient interventions can improve knowledge, reasoning, and clinical performance when compared with no intervention or traditional formats, although effects depend on instructional design \cite{cook_computerized_2010}. Representation scaffolds can improve diagnostic efficiency by helping medical students process information more selectively \cite{braun_representation_2017}. More recent work on AI-generated feedback after virtual-patient encounters suggests that automated feedback may support medical history taking, communication, and OSCE-based performance \cite{borg_ai-generated_2026}. Together, these studies indicate that the educational value of digital technologies lies not merely in efficiency, but in their ability to support standardized practice, timely feedback, and learner self-regulation.

Large language models (LLMs) and LLM-based agents offer new possibilities for scalable simulated patient systems. Early work suggested that, if the medical knowledge and reasoning performance of models such as ChatGPT are sufficiently reliable, conversational interfaces could serve as simulated patients, feedback tools, or small-group learning partners \cite{gilson2023does}. Reviews have also shown that artificial intelligence and generative AI are rapidly entering learning support, assessment, feedback, and simulation-based teaching in medical education, while many tools remain at an early stage with heterogeneous designs and evaluation evidence \cite{gordon2024scoping, preiksaitis2023opportunities}. For the present study, the central issue is not whether AI should be added to curricula in general, but whether LLM agents can be organized around specific instructional functions that scaffold simulated clinical interviewing without substituting for students' own reasoning.

This distinction is important because generative AI in medical learning creates both opportunities and risks. Retrieval-augmented teaching assistants can constrain model responses to course materials and reduce hallucination risk \cite{thesen2025generative}, yet student use of GenAI can also lead to over-reliance, with reported associations between GenAI dependence and weaker critical thinking \cite{fu2026mixed}. The risk of AI-induced ``never-skilling'' has recently been articulated as the possibility that trainees who rely on AI too early may fail to develop the independent reasoning required for safe clinical practice \cite{ke_ai-induced_2026}. Accordingly, AI systems for early clinical training should scaffold rather than replace student reasoning: prompts should guide reflection, expose missing information, and support patient-centered communication without directly giving diagnostic answers.

Current research on LLM-based simulated patients has primarily focused on whether such systems can generate realistic and coherent patient interactions \cite{liu_development_2025, weisman_development_2025, li_large_2026}. Although fidelity is an important first step, conversational plausibility alone is insufficient to constitute an effective learning environment \cite{hamstra_reconsidering_2014, pico_realism_2026}. Students need not only to interact with a realistic patient but also to elicit relevant details, identify missing information, communicate empathically, compare diagnostic possibilities, and receive feedback that supports reflection \cite{ende_feedback_1983, fuentes-cimma_designing_2024, spooner_risky_2024}. Thus, the design focus has shifted from whether LLMs can imitate patients to how LLM-based agents can be organized to improve the process quality of simulated clinical interviews.

\begin{figure}[!t]
    \centering
    \includegraphics[width=1\linewidth]{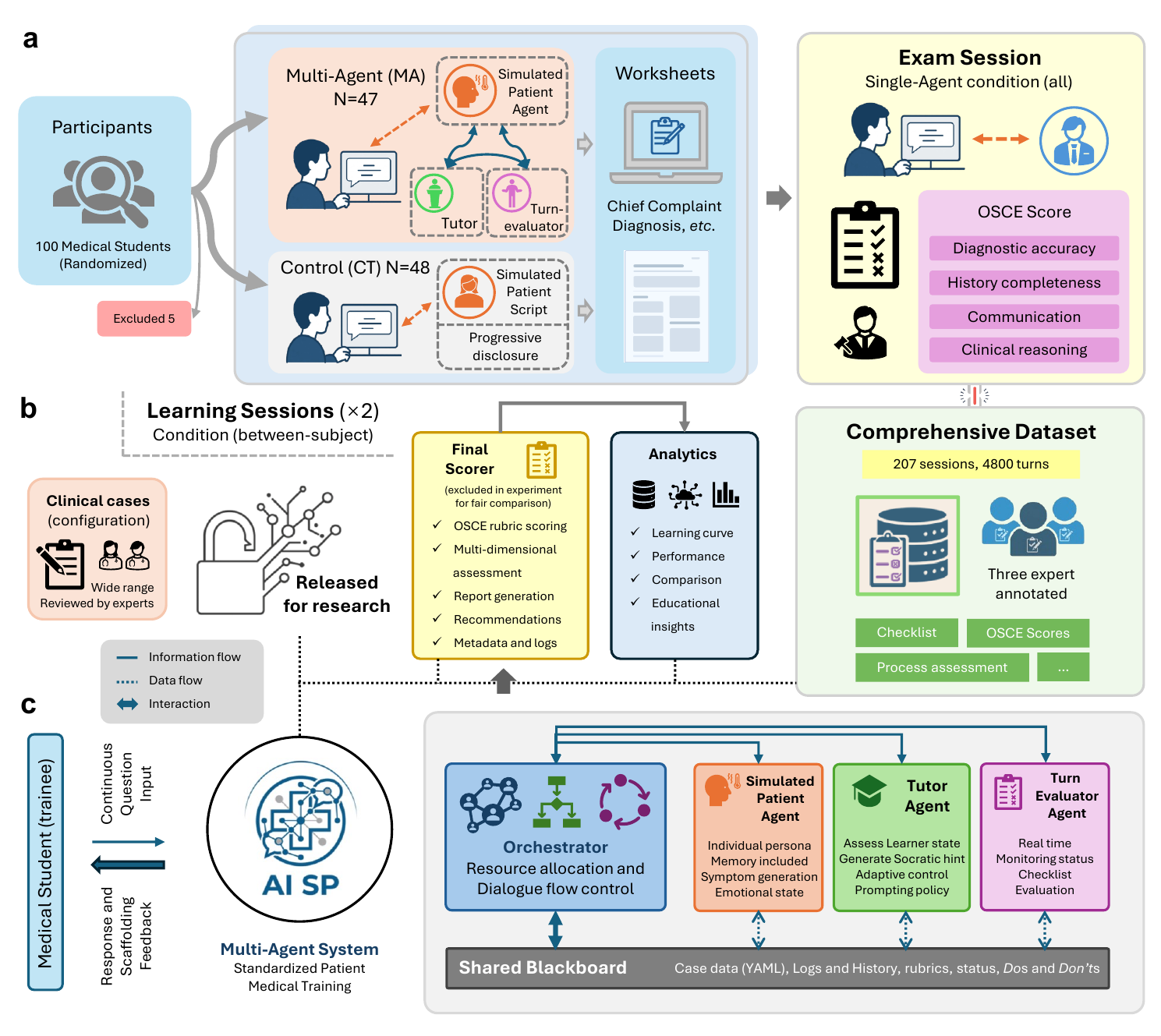}
    \caption{\textbf{Overall experimental framework and system architecture of MeduAI-SP.}
\textbf{(a)} Experimental design, consisting of two learning sessions and one exam session. Students were randomized to the multi-agent scaffolding condition (MA) or the structured non-LLM progressive-disclosure control condition (CT).
\textbf{(b)} Analysis pipeline and data open-source map.
\textbf{(c)} Architecture of the multi-agent system, illustrating the overall structure of the standardized patient medical training framework.}
    \label{fig:main}
\end{figure}

Multi-agent architectures offer a feasible approach to this instructional design challenge because different educational functions can be assigned to distinct specialized agents. In the broader higher education literature, multi-agent AI systems have been used to simulate instructional roles such as teachers, teaching assistants, and peers, with reported effects on knowledge construction, co-regulation, learning outcomes, and post-learning motivation, particularly among students with lower prior knowledge \cite{hao_mapping_2026}. In the medical AI context, agent-based workflows have also shown how LLM agents can be organized into specialized roles \cite{yu_simulated_2025}. Nevertheless, architecture alone does not constitute pedagogy. Existing multi-agent simulated patient systems have primarily demonstrated technical feasibility, task decomposition, or interface credibility \cite{du_llms_2024, zeng_embracing_2025}; empirical evidence remains limited on whether organizing LLM agents around explicit scaffolding functions improves simulated clinical interviewing compared with structured non-LLM learning materials derived from the same cases.

Scaffolding theory and the zone of proximal development (ZPD) provide the theoretical basis for this design. Students may not yet be able to conduct a complete, empathic, and diagnostically focused clinical interview independently, but they may be able to do so with timely prompts, feedback, and guided reflection \cite{kantar_rethinking_2020, masava_scaffolding_2022}. Medical education studies similarly support the role of scaffolding in diagnostic reasoning, although evidence is not uniformly positive \cite{braun_scaffolding_2019, nelson_use_2024}. Cognitive load theory further suggests that scaffolding should be timely, structured, and phased, reducing extraneous cognitive load without replacing learners' own reasoning processes \cite{leppink_evolution_2015, si_using_2024, young_cognitive_2014}.

Building on these foundations, we designed MeduAI-SP as an LLM-based multi-agent learning environment for simulated clinical interviewing. During history taking, the system integrates multiple instructional functions: an AI standardized patient provides inquiry-dependent case responses, a teaching agent provides timely Socratic prompts without revealing diagnostic answers, a turn-based evaluator monitors whether key history-taking and communication items have been addressed, and a final evaluator summarizes performance after the encounter. The tutor prompts were designed to cover focused history taking, diagnostic reasoning, summarizing and confirming information, and empathic or rapport-building communication.

We evaluated the platform through a randomized controlled study involving 100 medical students (Figure~\ref{fig:main}a). Participants were randomly assigned to either the structured non-LLM control condition or the multi-agent scaffolding condition. Each student completed two learning encounters under the assigned condition, followed by one examination encounter with a patient-only AI standardized patient, ensuring consistency in the testing environment. Examination performance was assessed using an Objective Structured Clinical Examination (OSCE)-aligned framework \cite{harden_assessment_1975}, including an overall score, domain-specific ratings, checklist-based behavior coverage, and diagnostic accuracy.

We formulated two primary hypotheses and several exploratory questions. First, we hypothesized that students in the multi-agent scaffolding condition would achieve higher final OSCE-aligned examination performance than students in the structured control condition. Second, we hypothesized that the strongest benefits would appear in observable consultation behaviors---especially patient-centered communication, empathic expression, and structured history taking---rather than necessarily in binary diagnostic accuracy, because both groups received the same core case content. Exploratory analyses examined learning trajectories, multidimensional performance phenotypes, process measures, survey responses, and the reliability of expert annotations.

In addition to the intervention study, we curated and released an open dataset derived from authentic instructional use cases on the platform. The dataset contains 207 clinical history-taking sessions and 4,815 turns of human-AI dialog interactions between medical students and AI standardized patients. Three clinical experts annotated the dataset using a multidimensional framework for clinical history-taking competence, supporting evaluation of AI-SP fidelity, student dialogue behavior, scaffolding needs, and OSCE-aligned performance. By releasing the dataset, annotation framework, and baseline models, we aim to advance future research on scalable and trustworthy AI standardized patients and automated assessment in medical education.

\section{Results}

\begin{figure}[htbp]
    \centering
    \includegraphics[width=1\linewidth]{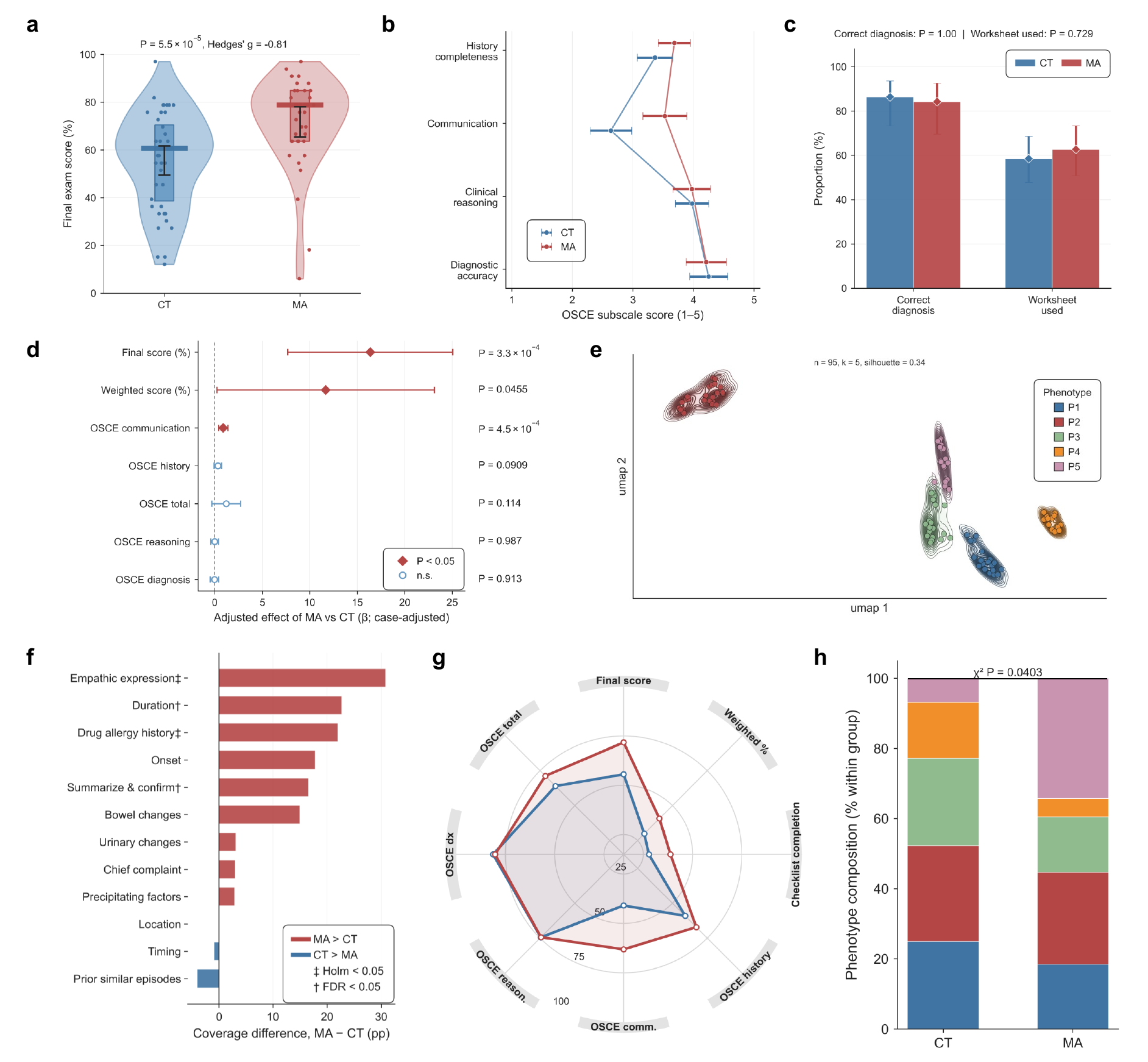}
    \vspace{0.5em}
    \caption{\textbf{Comparative exam-phase performance between the structured control group (CT; n = 48) and multi-agent dialogue group (MA; n = 47)}
\textbf{(a)} Session-level final exam scores (\%). Violin plots show score distributions; points indicate individual sessions; colored band denotes the median; box represents interquartile range (IQR); whiskers indicate mean $\pm$ 95\% CI.
\textbf{(b)} OSCE subscale scores (1--5) from the final evaluator shown as group means $\pm$ 95\% CI.
\textbf{(c)} Proportions of sessions achieving correct primary diagnosis and completed worksheet.
\textbf{(d)} Linear-model coefficients ($\beta \pm$ 95\% CI) for MA vs. CT. Filled diamonds indicate $P < 0.05$.
\textbf{(e)} Exploratory exam-performance phenotypes identified via UMAP embedding of standardized OSCE subscales and final score, followed by KMeans clustering ($k = 5$); Contours represent within-cluster density.
\textbf{(f)} Checklist item coverage difference (percentage points); symbols indicate multiplicity-adjusted significance (Holm or FDR).
\textbf{(g)} Multidimensional exam profile displayed on a common 0-100 scale.
\textbf{(h)} Distribution of identified phenotypes within CT and MA arms (\% within arm; exploratory comparison).
MA was associated with higher final exam scores and improved communication-related performance, while diagnostic accuracy and worksheet completion were comparable between arms. Exploratory analyses identified distinct performance phenotypes with differing distributions across arms.}
    \label{fig:mainresult}
\end{figure}

\subsection{MeduAI-SP platform implementation}
\label{sec:platform}

We implemented MeduAI-SP (\textbf{Med}ical \textbf{Edu}cation \textbf{A}rtificial \textbf{I}ntelligence--\textbf{S}tandardized \textbf{P}atient) as the intervention and assessment environment for this study. The platform supported the multi-agent learning condition, the structured non-LLM control condition, and a patient-only examination mode. Detailed information on the platform architecture, case materials, control condition, agent roles, prompt management, and data capture is provided in the Methods and Supplementary Information.

A total of 100 students were enrolled and randomly assigned to either the multi-agent AI standardized patient learning group (MA) or the structured control group (CT). Ninety-five students (MA, n = 47; CT, n = 48) completed the full experimental workflow and were included in the primary analysis.

\subsection{Multi-agent AI standardized patient training improved final examination performance}

Compared with students in the structured control group (CT, n = 48), students trained with the multi-agent AI standardized patient system (MA, n = 47) achieved higher final examination scores. The mean final examination score was \(71.8\%\) in the multi-agent group and \(55.6\%\) in the control group, with corresponding medians of \(78.8\%\) and \(60.6\%\), respectively. A Mann-Whitney U test showed a statistically significant between-group difference \((P = 5.51 \times 10^{-5})\), with a large standardized effect size favoring the multi-agent group \((\text{Hedges' } g = -0.81)\). The negative sign reflects the coding direction of the comparison rather than an unfavorable effect of the intervention (Figure~\ref{fig:mainresult}a).

The advantage of the multi-agent group was confirmed in a linear model. Assignment to the multi-agent group was associated with a 16.4-percentage-point higher final examination score compared with the control group \((\beta = 16.4;\ 95\%\ \mathrm{CI},\ 7.7\ \text{to}\ 25.1;\ P = 3.30 \times 10^{-4})\) (Figure~\ref{fig:mainresult}d). Multi-agent learning was also associated with a higher weighted checklist score \((\beta = 11.7;\ 95\%\ \mathrm{CI},\ 0.2\ \text{to}\ 23.1;\ P = 0.046)\), suggesting broader or more complete coverage of expected clinical behaviors during the final examination.

\subsection{Largest OSCE-domain improvement was observed in communication}

Analysis of OSCE domain scores showed that the most pronounced between-group difference occurred in communication. On the 1--5 OSCE rating scale, the mean communication score was higher in the multi-agent group than in the control group \((3.53\ \text{vs.}\ 2.64)\). This difference was statistically significant \((P = 4.14 \times 10^{-4})\) and corresponded to a large effect size \((g = -0.79)\), again with the sign reflecting the coding direction of the comparison (Figure~\ref{fig:mainresult}b).

Regression analyzes yielded similar results. Assignment to the multi-agent group was associated with a 0.90-point increase in the OSCE communication score \((\beta = 0.90;\ 95\%\ \mathrm{CI},\ 0.41\ \text{to}\ 1.39;\ P = 4.50 \times 10^{-4})\) (Figure~\ref{fig:mainresult}d). History-taking completeness showed a weaker positive association that did not reach conventional statistical significance \((\beta = 0.34;\ 95\%\ \mathrm{CI},\ -0.05\ \text{to}\ 0.73;\ P = 0.091)\). In contrast, descriptive differences in clinical reasoning scores were relatively small. See Supplementary Table~\ref{tab:s3_exam_outcomes} for details.

Together, these findings indicate that the most visible benefit of the multi-agent AI standardized patient system was not limited to checklist completion. Rather, the intervention appeared to have a particularly strong effect on communication-related OSCE performance. Multidimensional assessment profiles further supported this interpretation: the multi-agent group showed higher scores in communication and weighted checklist performance, whereas diagnosis-related dimensions showed substantial overlap between groups (Figure~\ref{fig:mainresult}g).

\subsection{Item-level analysis emphasizes empathetic expression rather than diagnosis }

Although the multi-agent group achieved higher final examination scores and better communication performance, the two groups did not differ in binary diagnostic accuracy. Correct initial diagnoses were made in \(86\%\) of control cases and \(84\%\) of multi-agent cases \((P = 1.000)\) (Figure~\ref{fig:mainresult}c). Worksheet completion also did not differ significantly between groups \((P = 0.729)\).

To further identify which clinical behaviors contributed to higher weighted checklist scores, we compared item-level checklist coverage between groups. The largest absolute difference was observed for the item \textit{expressing empathy}, which was 31 percentage points higher in the multi-agent group than in the control group (Figure~\ref{fig:mainresult}f). This difference remained statistically significant after Holm correction \((P = 8.30 \times 10^{-4})\). Because the comparator was a structured non-LLM progressive-disclosure condition rather than human standardized-patient feedback, this result is best interpreted as improved observable empathic communication behavior within the AI-SP examination setting.

Several other checklist items also showed favorable differences for the multi-agent group. The \textit{medication allergy history} item remained statistically significant after Holm correction, whereas \textit{duration of illness} and \textit{summarizing and confirming information} reached significance under false discovery rate \((\mathrm{FDR})\) correction. Additional items, including onset of symptoms, bowel habit changes, urinary changes, and chief-complaint inquiry, showed positive descriptive differences. These item-level findings were consistent with the OSCE-domain results, suggesting that the multi-agent intervention improved both communication style and selected components of structured history taking.

\subsection{Exploratory phenotyping revealed heterogeneous examination performance profiles}

We next explored whether final examination performance could be decomposed into distinct multidimensional phenotypes. Standardized OSCE domain scores and final examination scores were combined to generate a UMAP embedding, followed by \(k = 5\) K-means clustering. This exploratory analysis identified five performance phenotypes across the 95 completed examinations, with a silhouette score of 0.34 (Figure~\ref{fig:mainresult}e). The clusters occupied distinguishable regions in the embedding space, representing different combinations of OSCE-domain performance and overall final examination scores.

The distribution of phenotypes differed between the control and multi-agent groups \((\chi^{2} = 16.15;\ P = 0.040)\) (Figure~\ref{fig:mainresult}h). Because this analysis was exploratory and conducted post hoc, the findings should be interpreted as hypothesis-generating. Nevertheless, the phenotypic analysis suggests that the intervention may have shifted students toward different examination performance profiles rather than producing a uniform numerical improvement across all dimensions.

\subsection{Analysis of learning trajectories within the multi-agent group}

\begin{figure}[!t]
    \centering
    \includegraphics[width=1\linewidth]{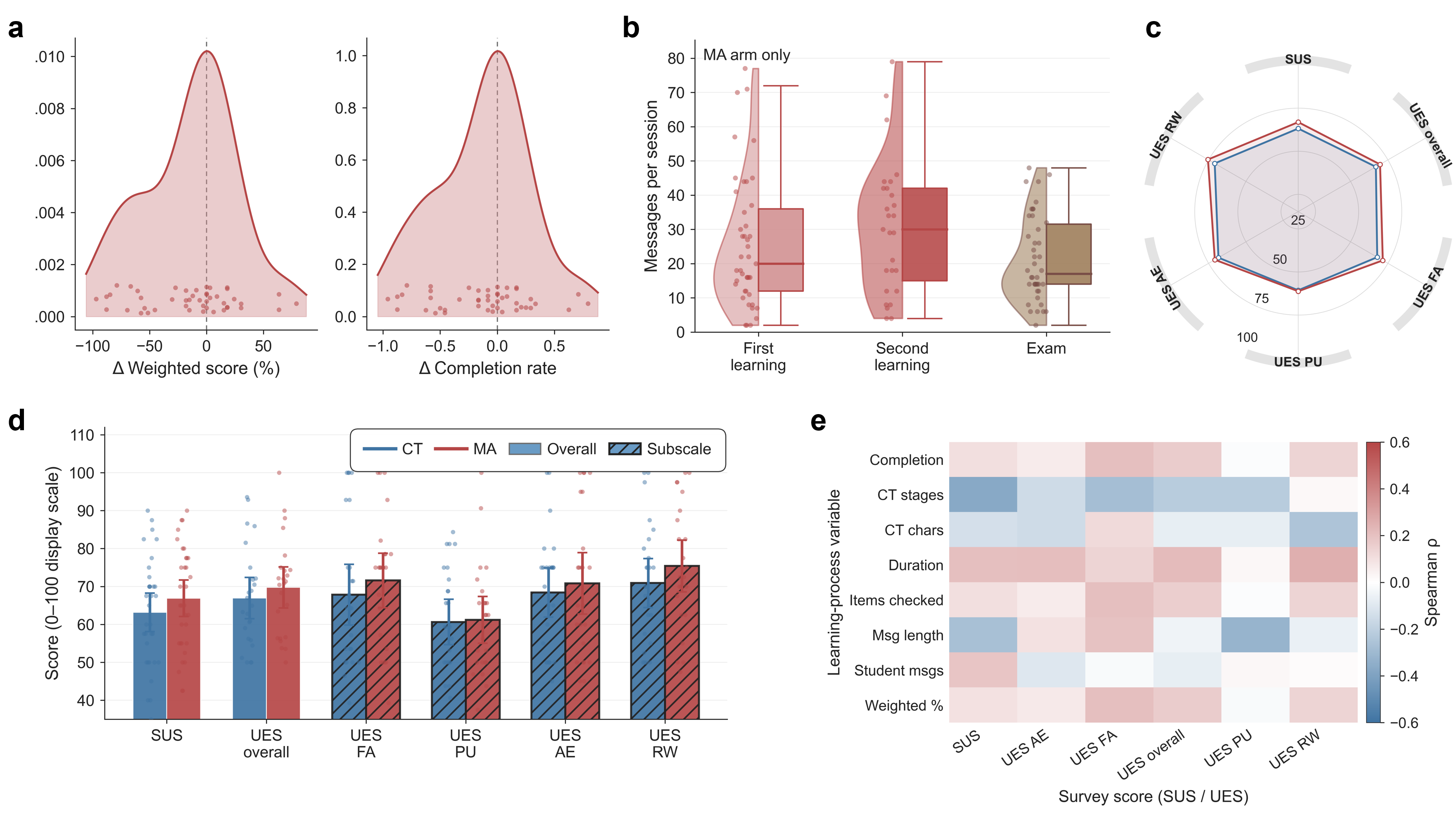}
    \caption{\textbf{Learning-phase performance, process measures and user experience outcomes.}
\textbf{(a)} Within-subject learning gain in the MA arm. Left: Change in weighted checklist score (percentage points; $\Delta = \text{session 2} - \text{session 1}$). Kernel density estimate with overlaid individual participants.  Right: Change in completion rate ($\Delta$ completion rate) in the MA arm.
\textbf{(b)} Messages per session in the MA arm across first learning, second learning and exam phases. The left half of each distribution shows a violin plot (session-level distribution), and the right half shows a box plot (inter-quartile range, median and whiskers); points represent individual sessions. MA arm only.
\textbf{(c)} Radar plot of survey experience profile on a common 0--100 display scale. System Usability Scale (SUS) scores and User Engagement Scale (UES) subscale scores were rescaled for display. Group means are shown as polygons for the CT and MA arms.
\textbf{(d)} Bar charts of all survey dimensions on a 0--100 display scale. Solid bars represent overall scores (SUS and UES overall), and black-hatched bars represent UES subscales: Focused Attention (FA), Perceived Usability (PU), Aesthetic Appeal (AE), and Reward Factor (RW). Bars indicate mean $\pm$ 95\% confidence interval; points denote individual participants.
\textbf{(e)} Exploratory Spearman correlations between learning-phase process variables (rows) and post-intervention SUS/UES scores (columns). Correlation coefficients ($\rho$) are displayed; analyses are exploratory and not preregistered primary endpoints.}
    \label{fig:process}
\end{figure}

Within the multi-agent group, we examined changes between the first and second learning sessions (Figure~\ref{fig:process}a). Changes in participant-level weighted checklist scores were defined as the score in the second learning session minus the score in the first learning session. These changes did not show a uniform positive increase. Among 46 participants, the mean change was \(-14.4\) percentage points, and the median change was \(-4.5\) percentage points. The corresponding change in completion rate was also slightly negative, with both the mean and median equal to \(-0.1\).

Visual inspection of the distribution of participant-level changes suggested heterogeneous learning trajectories rather than a single homogeneous response pattern. Specifically, the distribution suggested the possible presence of two learner subgroups. Because this analysis was not a prespecified endpoint, these findings should be interpreted descriptively.

We also quantified interaction volume in the multi-agent group by counting the number of messages exchanged during the first learning session, the second learning session, and the examination phase. The median numbers of messages were 20.0, 30.0, and 17.0, respectively (Figure~\ref{fig:process}b). Thus, students generated the greatest interaction volume during the second learning session, whereas message counts were lower during the examination phase. Compared with the first learning session and the examination phase, learners interacted more frequently during the second multi-agent learning session.

\subsection{Usability and Engagement Analysis}

Post-intervention survey data indicated a generally favorable user experience for both learning platforms. The mean System Usability Scale \((\mathrm{SUS})\) \cite{brooke1996sus} score was 63.2 in the control group and 66.9 in the multi-agent group, with a survey completion rate of approximately \(78\%\) (Figure~\ref{fig:process}c). User Engagement Scale \((\mathrm{UES})\) \cite{obrien_practical_2018} scores, rescaled to a 0--100 range, were also favorable in both groups. The mean total UES score was 69.8 in the multi-agent group and 66.9 in the control group, and these survey comparisons were treated as exploratory descriptive outcomes rather than confirmatory evidence of a group difference. The completion rate for the UES was approximately \(58\%\).

Radar plots of survey domains showed broadly similar experience profiles across the two platforms, with slightly higher mean values for the multi-agent platform in the focused-attention domain (Figure~\ref{fig:process}c). Distributions of SUS, total UES, and the four UES subscales---Focused Attention, Perceived Usability, Aesthetic Appeal, and Reward Factor---indicated broadly positive participant feedback in both groups (Figure~\ref{fig:process}d).

Among respondents with complete scale data, the survey instruments showed high internal consistency. The SUS demonstrated good reliability \((\text{Cronbach's } \alpha = 0.86)\), whereas the total UES showed excellent reliability \((\alpha = 0.96)\). Reliability was also high for the UES subscales, including focused attention \((\alpha = 0.95)\), perceived usability \((\alpha = 0.87)\), aesthetic appeal \((\alpha = 0.94)\), and reward \((\alpha = 0.95)\). These estimates support the internal consistency of the post-intervention questionnaire measures in this study.

Finally, we examined associations between process variables during the learning phase and post-intervention SUS and UES scores. These analyzes were not prespecified primary endpoints and should therefore be considered exploratory. The strongest observed associations included a negative correlation between the number of CT stages and SUS score \((\text{Spearman's } \rho = -0.37)\), a negative correlation between message length and UES perceived usability \((\rho = -0.32)\), and a negative correlation between the number of CT stages and UES focused attention \((\rho = -0.28)\) (Figure~\ref{fig:process}e). These associations suggest that longer or more complex learning processes did not necessarily translate into higher perceived usability or engagement, although the exploratory design precludes causal inference.

\begin{table}[!t]
\centering
\footnotesize
\setlength{\tabcolsep}{4pt}
\renewcommand{\arraystretch}{1.02}

\caption{\textbf{Dataset composition and expert annotation reliability.}}
\label{tab:dataset_reliability}

\begin{threeparttable}
\begin{tabularx}{\linewidth}{
>{\raggedright\arraybackslash}p{0.21\linewidth}
>{\raggedright\arraybackslash}p{0.34\linewidth}
>{\raggedright\arraybackslash}p{0.17\linewidth}
>{\raggedright\arraybackslash}X
}
\toprule
\textbf{Section} & \textbf{Item or construct} & \textbf{Statistic or unit} & \textbf{Value} \\
\midrule

\multicolumn{4}{l}{\textbf{Dataset composition}} \\

Sessions
& Annotated sessions
& Count
& 207 \\

Session type
& Multi-agent learning; examination
& Count
& 118; 89 \\



Messages
& Student; patient; tutor
& Count
& 2,350; 2,350; 115 \\

Messages/session
& Median; mean; range
& Messages
& 18; 23.3; 2--79 \\

Annotation units
& D1; D2/D3; D4
& Unit
& 2,350 patient messages; 2,350 student messages; 207 sessions \\

Annotators
& Expert-confirmed annotation tracks
& Gold standard
& Majority vote or mean score \\

\midrule
\multicolumn{4}{l}{\textbf{Inter-expert reliability}} \\

Dialogue act (D2)
& Eight-class dialogue-act label
& Fleiss' $\kappa$
& 0.84; almost perfect \\

OSCE domains (D4)
& History taking; clinical reasoning; communication; diagnosis
& ICC(2,$k$)
& 0.81; 0.80; 0.83; 0.85; good \\

Need for scaffolding (D3)
& Binary scaffolding-trigger label
& Fleiss' $\kappa$
& 0.56; moderate; 24.1\% trigger rate \\

Progressive disclosure (D1)
& Patient-message disclosure-control label
& Fleiss' $\kappa$
& 0.31; fair; 99.3\% pass rate \\

Script adherence (D1)
& Patient-message script-adherence label
& Fleiss' $\kappa$
& $\approx$0; degenerate; $\geq$99.8\% pass rate \\

\bottomrule
\end{tabularx}

\begin{tablenotes}[flushleft]
\footnotesize
\item Note: Categorical constructs were evaluated using Fleiss' $\kappa$, whereas OSCE domain scores were evaluated using ICC(2,$k$). Reliability constructs are ordered from highest to lowest agreement, with annotation dimensions indicated in parentheses. Landis--Koch descriptors are provided as qualitative references only. Script adherence was treated as degenerate because of the near-universal pass rate.
\end{tablenotes}
\end{threeparttable}
\end{table}

\subsection{Construction of an annotated consultation corpus}

\begin{figure}[htbp]
    \centering
    \includegraphics[width=1\linewidth]{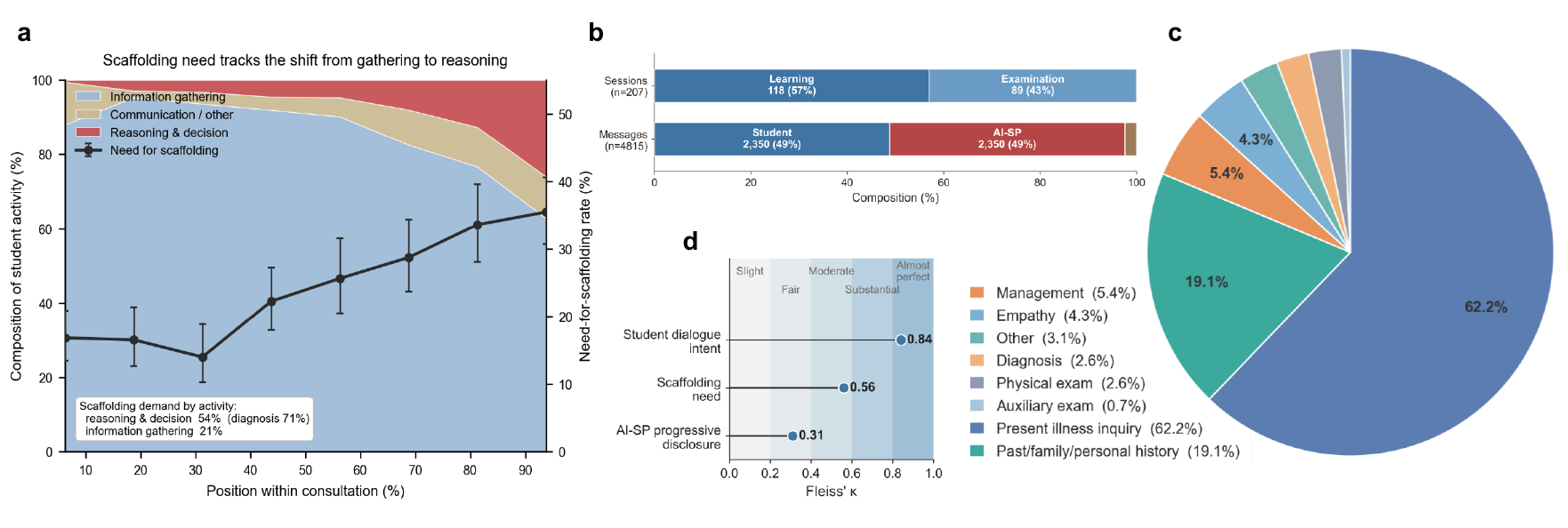}
    \caption{\textbf{Corpus composition, annotation reliability, student dialogue intent, and scaffolding dynamics across consultation.}
\textbf{(a)} Distribution of annotated student information categories and corresponding scaffolding rates. \textbf{(b) }Composition of the annotated corpus. \textbf{(c)} Distribution of student dialogue intent across 2,350 utterances, dominated by history-taking (62.2\% history of present illness; 19.1\% past/family/personal history). \textbf{(d)} Inter-rater agreement for annotation dimensions (Fleiss’ $\kappa$).
}
    \label{fig:dataset}
\end{figure}

To support process-level analysis of learner-patient interactions, we constructed an annotated consultation corpus from the multi-agent learning and examination sessions. The corpus contained 207 consultation sessions, including 118 multi-agent learning sessions and 89 examination sessions. Across all sessions, there were 4815 messages: 2350 student messages, 2350 AI standardized patient \((\mathrm{AI\mbox{-}SP})\) messages, and 115 tutor messages. The median number of messages per session was 18, and the median number of student utterances per session was 9 (Figure \ref{fig:dataset}b).

The annotation scheme was organized into four complementary dimensions, each corresponding to a different unit of analysis. First, all 2350 AI-SP messages were annotated for patient fidelity, including whether the response adhered to the case script and whether information was disclosed progressively rather than prematurely. Second, all 2350 student messages were annotated for dialog intent, including history taking, physical examination, ancillary testing, diagnosis, management, empathy or reassurance, and other utterance types. Third, the same 2350 student messages were annotated for whether instructional scaffolding from the tutor was needed, together with the reason for intervention and the recommended pedagogical strategy. Fourth, all 207 sessions were assigned global OSCE ratings at the consultation level. Thus, the corpus linked turn-level behavior, patient simulation quality, tutor-intervention opportunities, and session-level clinical performance within a single annotated dataset (Table~\ref{tab:dataset_reliability}).

\subsection{Student-patient interactions and scaffolding needs}

Turn-level annotations showed that the consultations were primarily organized around history taking, particularly the history of present illness (Figure \ref{fig:dataset}c). Among student utterances, questions about the history of present illness accounted for 62.2\%, whereas questions about past medical history, family history, and personal history accounted for 19.1\%. Other utterance intents were less frequent, including management or treatment discussions (5.4\%), empathic or supportive communication (4.3\%), other utterances (3.1\%), diagnostic statements (2.6\%), physical examination-related utterances (2.6\%), and ancillary-test-related utterances (0.7\%). The most common information targets included associated gastrointestinal symptoms, pain location, past medical history, precipitating factors, fever or systemic symptoms, and pain characteristics, consistent with the clinical priorities of acute abdominal pain assessment.

Expert annotations further indicated that approximately 24.1\% of student utterances required instructional scaffolding. The need for scaffolding was not evenly distributed across the encounter. It increased from approximately 16.1\% in the early phase of the consultation to approximately 34.8\% in the late phase, suggesting that learners more often required support when the encounter moved from initial information gathering toward integration, diagnostic reasoning, and management decisions (Figure \ref{fig:dataset}a). Among utterances requiring intervention, the most frequent reasons were conversational impasse, domain-specific knowledge gaps, communication breakdown, and premature closure (Section \ref{app:examples}).

\subsection{AI standardized patient responses showed high fidelity}

Annotations of AI-SP behavior showed that the simulated patient environment was stable across learning and examination sessions. Only approximately 0.68\% of AI-SP responses were judged to contain clear fidelity problems. Script adherence was almost uniformly acceptable, and progressive disclosure was rated as appropriate in approximately 99.3\% of patient messages. These findings suggest that the AI-SP generally maintained case consistency and disclosed information in a controlled, clinically plausible manner, providing a reliable simulation context for evaluating learner behavior and performance.

\subsection{Inter-rater agreement of expert annotations}

Inter-rater agreement varied across annotation dimensions (Table~\ref{tab:dataset_reliability}). Agreement was highest for student dialog intent, with a Fleiss' \(\kappa\) of 0.84, indicating near-perfect agreement. Agreement for global OSCE scoring was also good. The intraclass correlation coefficients (ICCs) for the mean rating of three experts, ICC\((2,k)\), ranged from 0.80 to 0.85 across the four OSCE domains, with the highest agreement for diagnostic accuracy \((\mathrm{ICC}(2,k)=0.85)\) and the lowest for clinical reasoning \((\mathrm{ICC}(2,k)=0.80)\). By contrast, single-rater reliability, ICC\((2,1)\), ranged from 0.58 to 0.66, supporting the use of the average expert rating as a more stable reference standard for OSCE-domain evaluation. Questionnaire internal consistency is reported separately in Supplementary Table~\ref{tab:s1_reliability.}

Agreement for whether a student utterance required scaffolding was moderate, with a Fleiss' \(\kappa\) of 0.56. Agreement was lower for AI-SP progressive disclosure \((\kappa=0.31)\), likely reflecting the highly imbalanced distribution of labels, as nearly all patient responses were rated as appropriate. Script adherence showed near-zero agreement because almost all responses were classified as adherent, leaving little between-rater variability to estimate. 

Overall, annotation dimensions based on directly observable utterance-level features and well-defined label sets showed stronger agreement, whereas dimensions involving highly imbalanced labels showed lower agreement. Nevertheless, the OSCE ratings demonstrated sufficient reliability for use in downstream performance analyzes.
\section{Discussion}

This randomized study evaluated MeduAI-SP, a web-based multi-agent AI standardized patient platform designed to support medical students' simulated consultation practice. The primary finding was that multi-agent AI standardized patient training improved final examination performance compared with a structured non-LLM progressive-disclosure control condition. Students in the multi-agent group achieved higher final examination scores and higher weighted checklist scores. The largest domain-level improvement was observed in OSCE-style communication performance, and item-level analyzes showed particularly large gains in observable empathic expression, medication allergy history, summarizing or confirming information, and selected history-taking behaviors.

In contrast, binary diagnostic accuracy did not differ between groups. This result was expected given the study design. Both groups were exposed to the same authored clinical cases and the same core diagnostic content, and the multi-agent intervention was not intended to provide additional disease knowledge beyond the controlled case materials. Therefore, the absence of a diagnostic accuracy difference should not be interpreted as a failure of the intervention. Rather, it suggests that the experimental control was appropriate: the observed benefit of the multi-agent system was not due to unequal access to diagnostic knowledge, but to differences in how learners conducted the consultation.

The strongest effect of the multi-agent system was observed in communication-related performance. The multi-agent group showed substantially higher OSCE communication scores, and the largest checklist difference was for empathic expression. These findings should be interpreted as an improvement in observable empathic communication behavior during an AI-SP simulated encounter, not as evidence that students became more empathetic as a stable personal attribute. They also should not be interpreted as evidence that AI-SP feedback is superior to, or can replace, human standardized-patient or faculty feedback on empathic communication, because no such comparator arm was included. Artificial intelligence and humans may make different contributions: AI may help maintain consistency and provide repetitive practice, while human educators and clinicians remain essential in interpreting complex emotional cues, adapting to social contexts, exercising judgment, and taking responsibility for consequential decisions.

This distinction is important in clinical education. Consultation competence is not limited to reaching the correct diagnosis; it also includes the observable behaviors through which clinical reasoning is enacted. Frameworks such as the Calgary-Cambridge guide describe consultation skills as teachable behaviors, including initiating the session, gathering information, building the relationship, explaining and planning, and closing the encounter \cite{kurtz_calgarycambridge_1996}. Similarly, Miller’s framework emphasizes that clinical competence includes what learners can show in structured performance settings, not only what they know or whether they can state the correct answer \cite{miller1990assessment}. The OSCE tradition was developed for this reason: to assess observable clinical performance using structured encounters and explicit criteria \cite{harden_assessment_1975}.

From this perspective, the present findings suggest that the multi-agent platform improved the conduct of the simulated consultation. Students trained with the multi-agent system were more likely to display patient-centered communication, elicit relevant information, and complete selected consultation behaviors. These process-level improvements are educationally meaningful.

Why might multi-agent scaffolding have improved consultation behavior? A plausible explanation is that the multi-agent design helped students manage the process demands of the consultation. The tutor and turn evaluator provided phased support for focused history taking, diagnostic reasoning, summarization, and empathic communication rather than relying on a single undifferentiated conversational agent. The design also intentionally avoided direct disclosure of diagnostic answers, aligning with the principle that AI should scaffold rather than substitute student reasoning.

The findings also suggest that early gains in clinical reasoning education may first appear in the behaviors that make reasoning possible. Before learners show measurable improvement in diagnostic endpoints, they may improve in asking clinically useful questions, confirming information, avoiding premature closure, and maintaining patient-centered communication. These behaviors are not substitutes for diagnostic accuracy, but they are part of the pathway through which diagnostic reasoning is performed in real consultations.

The present findings are most consistent with a model of functional complementarity. In medical education, AI and human educators need not perform identical forms of instructional or communicative work. AI agents may be especially useful for repetitive practice, consistent role-play, checklist monitoring, Socratic prompting, and the generation of preliminary formative feedback. Faculty members and human standardized patients may contribute contextual interpretation, nuanced emotional response, individualized remediation, assessment of professionalism, and judgments about learner readiness.

On this account, partial substitution may be acceptable at the level of specific and bounded tasks without implying replacement of the educator, standardized patient, or clinician as a whole. For example, an AI-SP may substitute for some repetitive early-stage interview practice, and an evaluator agent may provide preliminary feedback on checklist coverage. However, the system should not autonomously determine whether a learner is clinically competent, make progression or credentialing decisions, or replace human-led remediation in emotionally or professionally complex cases. The appropriate boundary depends on the consequences of error, the uncertainty of the task, the relational sensitivity of the interaction, and the availability of meaningful human review.

The present study has implications for how AI standardized patient systems should be evaluated. If evaluation focuses only on diagnostic accuracy, it may miss important educational effects. This pattern indicates that AI-supported simulated patient systems may influence how learners conduct the encounter before they improve terminal diagnostic accuracy. At the same time, process improvements should be evaluated alongside safeguards against over-reliance and AI-induced never-skilling, particularly for novice learners who still need to develop independent reasoning \cite{ke_ai-induced_2026}.

Patient perspectives are central to the evaluation of empathic communication, but they were not directly represented in the present experiment. The simulated patient was generated by an LLM, and student performance was assessed through checklist and OSCE-aligned measures. These measures capture observable behaviors that are educationally relevant, but they cannot establish whether a real patient would feel heard, understood, respected, reassured, or supported.

Emerging evidence suggests that patient evaluations of AI-generated communication are heterogeneous. In one study, people with cancer rated chatbot responses as more empathic than physician responses, supporting the possibility that AI-generated language can meet some patients’ communication needs \cite{chen2025patient}. In another study of asynchronous oncology communication, medical reviewers rated chatbot responses as more empathic, whereas the patient panel consistently favored physician responses \cite{bai2025application}. Perceived authorship may also influence ratings: responses can receive higher ratings when participants believe they were written by a physician, even when the actual text was AI-generated \cite{ruben2025artificial}. These findings suggest that patient preference is not uniform and may depend on clinical complexity, communication purpose, prior relationships, trust, and individual attitudes toward AI.

Patients’ felt experience of being understood should therefore carry substantial weight in evaluating AI-supported communication. However, perceived empathy should not be the sole criterion. It should be considered together with factual accuracy, clinical safety, the appropriateness of reassurance, cultural and linguistic responsiveness, privacy, equity, and access to human escalation.

Future evaluations of LLM-based simulated patient systems should therefore include both endpoint and process-sensitive outcomes. Diagnostic correctness remains important, but it should be complemented by measures of communication quality, history-taking completeness, information elicitation, checklist coverage, and consultation structure. This is especially important in early clinical training, where students are still learning how to conduct the consultation itself.

An additional contribution of this study is the construction of an annotated consultation corpus which provides a process-level view of how students interacted with simulated patients and where instructional support was needed.
The annotation results showed that student consultations were dominated by history taking, especially the history of present illness. This is appropriate for acute abdominal presentations, where symptom characterization and associated features are central to diagnostic reasoning. However, other utterance types were relatively infrequent, including physical examination, ancillary testing, diagnostic statements, management discussions, and empathic communication. These findings suggest that novice learners may focus heavily on gathering symptom information while underusing other components of a complete clinical consultation. These findings can inform future tutor-agent design by identifying when and why learners most often require support.

The dataset also provides evidence that the AI standardized patient environment was stable. Only a small proportion of AI-SP responses were judged to have clear fidelity problems, and progressive disclosure was rated as appropriate in almost all patient messages. This is important because educational interpretation depends on the reliability of the simulated patient.


\paragraph{Limitations.}

This study has several limitations. First, the comparator was a structured non-LLM progressive-disclosure control condition rather than a human standardized-patient or faculty-feedback arm. Therefore, the results support the benefit of multi-agent scaffolding relative to structured case materials, but they do not establish that AI-SP feedback is equivalent or superior to human feedback, nor do they justify replacing human feedback on empathic communication. Future studies should compare human SP plus faculty feedback, AI-SP plus faculty oversight, and hybrid models using the same outcomes.
Second, although the primary outcome was prespecified, the study was conducted within a limited clinical domain and with a modest sample size. The cases focused on acute abdominal or gastrointestinal presentations, and the primary examination analysis centered on a small number of authored cases. Accordingly, the findings should be generalized with caution to other specialties or authentic clinical environments.
Third, several analyzes were exploratory, including the phenotype analysis, learning-trajectory analysis within the multi-agent group, and correlations between process variables and post-intervention survey scores. These analyzes are informative for hypothesis generation but should not be considered confirmatory evidence.
Fourth, no real patients or human standardized patients participated in the intervention or primary examination. Consequently, the study cannot determine whether the observed communication behaviors would cause patients to feel heard, understood, respected, or supported. Expert-rated and checklist-based empathic expression should not be assumed to be equivalent to patient-perceived relational empathy. Future validation should include human standardized patients, patient-reported communication outcomes, and participants with diverse cultural, linguistic, demographic, and clinical backgrounds.
Fifth, the study did not include delayed follow-up. Consequently, it remains unclear whether the observed improvements in simulated consultation performance are sustained over time or transferred to higher-stakes clinical settings. Future work should incorporate delayed transfer assessments, faded scaffolding, and faculty oversight to test whether learners internalize independent interviewing and reasoning skills rather than becoming dependent on AI prompts.



\paragraph{Conclusion.}

MeduAI-SP, a web-based multi-agent AI standardized patient platform, improved prespecified final examination performance compared with a structured non-LLM control condition. The strongest gains were observed in communication and selected consultation behaviors, including observable empathic expression and information elicitation. The study also produced an annotated consultation corpus that characterizes learner dialogue, scaffolding needs, AI-SP fidelity, and OSCE-level performance. These data provide a foundation for future validation of automated assessment and for the design of more targeted tutor-agent interventions. Future work should compare AI-SP scaffolding with human standardized-patient and faculty-feedback conditions, test transfer to clinical settings, evaluate delayed retention, and examine how faded scaffolding can preserve independent reasoning while maintaining scalable practice opportunities.


\section{Ethics Approval and Consent}

This study was reviewed and approved by the Ethics Committee of Guangzhou Medical University
(Approval No. 202605032). All study procedures were conducted in accordance with relevant medical
ethics requirements, institutional regulations, and applicable ethical guidelines. At the beginning of
the experiment, all student participants were presented with an online informed consent form and were
required to confirm their consent before participation. Data collection was conducted under
strict privacy protection procedures, and all exported data were de-identified before analysis.

\section{Author Contributions}

L.Y. contributed to conceptualization, methodology, model development, experimental implementation, data analysis, and manuscript preparation. 
H.L. contributed to methodology, data curation, data collection, annotation, and formal analysis. 
S.L. contributed to formal analysis, visualization, model development, and writing the manuscript. 
R.J. contributed to writing the manuscript, editing, and provided educational expertise. 
Y.X. provided medical domain expertise and clinical consultation. 
G.C. contributed to supervision and technical support for large language model development. 
L.L. contributed to conceptualization, supervision, project administration, and data collection.
All authors reviewed and approved the final manuscript.



\section{Competing Interests}

The authors declare no competing interests.

\section{Funding}

Not Applicable.

\clearpage

\begin{spacing}{1.1}

\section{\texorpdfstring{Methods}{Methods}}

\subsection{Study design and cohort}

This study used a two-arm parallel randomized controlled design to evaluate the effectiveness of a large language model (LLM)-based multi-agent scaffolded learning environment for simulated clinical interviewing. Participants were randomly assigned to either the multi-agent learning group (MA group) or the structured control group (CT group). Both groups followed the same standardized experimental pipeline consisting of two learning encounters, one examination encounter, and a post-intervention questionnaire. The primary outcome was final OSCE-aligned examination performance. Secondary outcomes included OSCE-domain scores, weighted checklist coverage, diagnostic accuracy, worksheet completion, and post-intervention usability and engagement measures. Learning-trajectory, phenotype, process-correlation, and dataset-reliability analyses were exploratory.

Participants were third-year undergraduate students majoring in clinical medicine at Guangzhou Medical University. This cohort was selected because students at this stage are approaching formal standardized-patient-based clinical training and are beginning to develop clinical history-taking and reasoning skills. Students were recruited on a voluntary basis. Participants were randomized in a 1:1 ratio using a computer-generated random sequence. See Supplementary Table~\ref{tab:participants} for details.

\subsection{Materials}

The clinical cases used in the experiment were developed by a medical professor and independently reviewed by another medical professor. The platform contained an expanded library of 18 standardized clinical cases, covering typical emergency conditions such as acute appendicitis and acute pancreatitis, as well as diseases from other clinical systems.

For the present experiment, three acute abdominal cases were used: acute appendicitis presenting with migratory right lower-quadrant abdominal pain, acute pancreatitis presenting with epigastric pain with belt-like radiation, and perforated peptic ulcer presenting with sudden board-like abdominal rigidity (Section~\ref{supcasemat}). The first two cases were used as learning cases, and the third case was used as the examination case. All cases were represented using a unified YAML-based data structure, which allowed the same case content to be used across different experimental modes and ensured alignment of case difficulty and medical knowledge across conditions.

\begin{figure}[!t]
    \centering
    \includegraphics[width=0.9\linewidth]{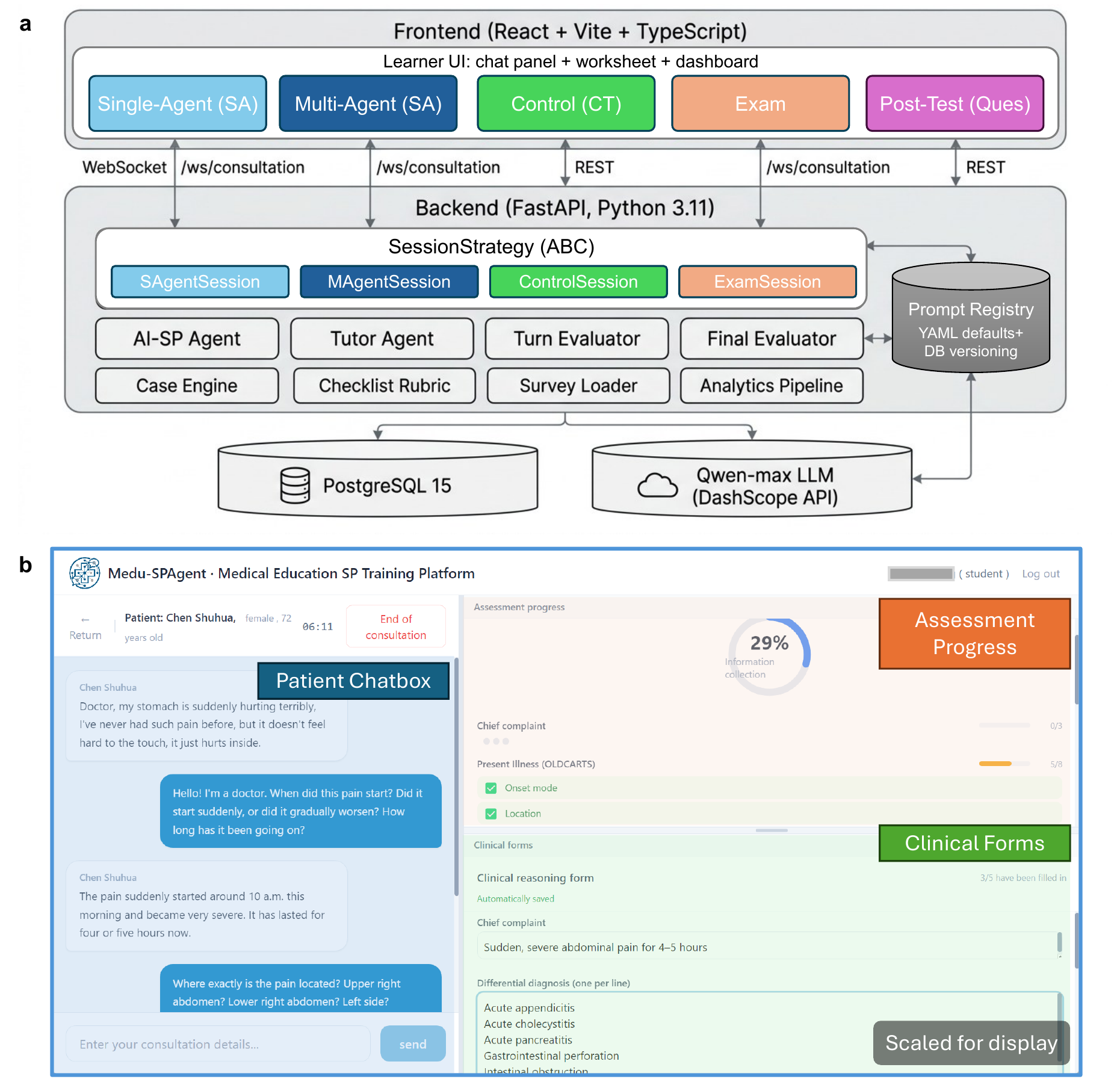}
    \caption{\textbf{AI Standardized Patient (SP) Training and Research Platform Architecture.}
\textbf{(a)} Frontend--backend--SQL architecture diagram illustrating the overall system design, including scalable and modular components that support extensibility and integration.
\textbf{(b)} Screenshot of the platform's multi-agent experimental web interface, featuring the patient chatbox, assessment progress panel, and clinical forms. Socratic prompts are displayed above the chatbox to guide structured history taking, clinical reasoning, and empathic communication.}
    \label{fig:methodplatform}
\end{figure}

\subsection{Platform design}

The experiment was conducted on MeduAI-SP, an AI standardized patient training and research platform, as described in Section~\ref{sec:platform}. The platform supported four modes: single-agent learning, multi-agent learning, structured control learning, and examination. The present randomized comparison used the multi-agent learning mode as the intervention, the structured control learning mode as the comparator, and the examination mode as the common testing environment. After login and onboarding, each participant completed two learning sessions under the assigned condition. Both groups then completed one examination encounter with a new but difficulty-matched case. No real-time instructional feedback was provided during the examination.

The platform automatically recorded all interaction and assessment data in a PostgreSQL database, including user information, training sessions, dialogue messages, evaluation snapshots, final evaluations, control learning steps, survey responses, and prompt versions. Prompt versions and activation states were recorded in the database, and the currently active prompt versions were snapshotted at the beginning of each training session to support reproducibility.

\subsection{Multi-agent learning condition}

In the MA condition, students interacted with an AI standardized patient, a teaching agent, a turn-based evaluation agent, and a final evaluation agent. The patient agent generated inquiry-dependent responses based on the structured case script and was instructed to avoid premature disclosure of diagnostic clues. The teaching agent provided Socratic prompts during learning sessions. The turn-based evaluator monitored whether the student had covered key history-taking and communication items and whether the student appeared stuck, missed critical information, communicated in a way that impaired rapport, or made a potentially unsafe inference. The final evaluator generated a post-encounter summary and OSCE-aligned assessment, but it was not visible during the live examination interaction.

The tutor prompts were designed to scaffold both clinical reasoning and empathic communication. They covered four broad areas: focused history taking, diagnostic reasoning, summarizing or confirming information, and rapport-building communication. Tutor prompts were triggered only when the turn evaluator identified a meaningful need for intervention, such as missing key information, conversational impasse, premature closure, communication breakdown, or professionalism concerns. The tutor was instructed not to directly reveal the final diagnosis, key differential diagnoses, or case answers. Instead, prompts were phrased as reflective questions or brief hints, such as asking whether the current information was sufficient to support a conclusion, reminding the student to address the patient's expressed distress before continuing, or suggesting that the student summarize what had already been learned. Additional prompt examples are provided in Supplementary Table~\ref{tab:s6_prompt_examples.}

\subsection{Structured control condition}

The CT condition was a structured non-LLM control rather than a human standardized-patient or AI-dialogue comparator. Students did not converse freely with an AI patient or a human simulated patient. Instead, they completed a computer-based progressive-disclosure learning activity built from the same authored case materials used in the MA condition. Case information was displayed in fixed stages: students first saw a brief initial presentation, then proceeded through staged information disclosure with predetermined answers and case details. This design preserved exposure to the same core diagnostic content while removing LLM dialogue, tutor scaffolding, real-time process monitoring, and automated feedback.

The CT group used the same learning cases, examination case, time allocation, post-intervention survey, and assessment method as the MA group. However, the CT group did not receive real-time Socratic prompts, tutor feedback, AI-generated patient dialogue, or post-encounter formative feedback during the two learning encounters. Students recorded diagnostic reasoning elements, including diagnosis, rationale, and management suggestions, but no correctness feedback, scoring, or standard-answer feedback was provided before the examination.

\subsection{Examination condition}

The examination mode was identical for both groups. Students interacted with a patient-only AI standardized patient for the examination case. Teaching and process-monitoring agents were removed from the student-facing interface, and no Socratic prompts or real-time feedback were provided. The final evaluator assessed performance in the background after the encounter using the OSCE-aligned framework described below.

\subsection{OSCE-aligned assessment framework}

The primary outcome was students' clinical interviewing performance in the examination mode. Performance was assessed using an OSCE-aligned framework \cite{harden_assessment_1975} that combined an overall final examination score, four global domain ratings, and checklist-based behavior coverage. The four global domains were history taking, clinical thinking, communication and empathy, and diagnostic accuracy, each rated on a 1--5 scale. The shared anchors were: 1 = clearly insufficient, 2 = poor, 3 = meets basic expectations, 4 = good, and 5 = excellent. Domain-specific descriptors are provided in Supplementary Table~\ref{tab:s5_osce_rubric.}

The weighted checklist captured observable consultation behaviors emphasized in the case scripts and annotation manual. Items included opening and communication behaviors (self-introduction, open-ended opening, empathy or comfort, summarization or confirmation), OLDCARTS-based pain characterization, associated gastrointestinal, systemic, urinary, and gynecologic symptoms, past medical, surgical, medication, allergy, family, and social history, physical examination, laboratory or imaging plans, diagnosis and differential diagnosis, and management planning. Expressing empathy was represented both as an item-level observable behavior and as part of the global communication-and-empathy domain. Case-specific alignment of key history items, expected diagnosis, and differential diagnoses is provided in Supplementary Table~\ref{tab:s7_case_alignment}.

Final diagnostic accuracy was coded as correct or incorrect according to whether the final primary diagnosis matched the expected diagnosis for the examination case. Checklist completion rate was defined as the proportion of predefined key items successfully covered by the student. Because the scoring framework was implemented from the study-specific case checklist and OSCE-style global ratings, we describe the observed scoring outputs rather than imposing an additional externally derived weighting formula not used in the original analysis.

\subsection{Post-intervention survey instruments}

After the examination, participants completed post-intervention questionnaires assessing usability and engagement. The manuscript uses the System Usability Scale (SUS) and the User Engagement Scale (UES); no separate Student Engagement Scale (SES) was administered. SUS is a 10-item usability instrument \cite{brooke1996sus}. Following standard scoring, positively worded items were scored as item response minus 1, negatively worded items were reverse-scored as 5 minus item response, adjusted item scores were summed, and the total was multiplied by 2.5 to yield a 0--100 score.

Engagement was measured using the 30-item User Engagement Scale \cite{obrien_practical_2018}, including four subscales used in the present analysis: Focused Attention, Perceived Usability, Aesthetic Appeal, and Reward Factor. UES responses were scored according to the instrument structure and rescaled to a common 0--100 display scale for visualization. Questionnaire completion differed by scale because some participants did not complete all survey items. Survey analyses used complete-case data for each scale or subscale; no mean imputation was applied. Cronbach's \(\alpha\) was used to assess internal consistency reliability for SUS, overall UES, and UES subscales (Supplementary Table~\ref{tab:s1_reliability}). Open-ended questionnaire responses were collected to supplement quantitative findings.

\subsection{Statistical analysis}

All statistical analyses were conducted using Python. Descriptive statistics were used to summarize participant characteristics, examination scores, diagnostic accuracy, checklist completion rates, interaction features, survey outcomes, and dataset properties. Continuous variables were summarized using means and standard deviations or medians and interquartile ranges, depending on distributional characteristics. Categorical variables were summarized using frequencies and percentages.

Between-group comparisons examined differences between the MA and CT groups in examination performance, diagnostic accuracy, checklist completion, and survey outcomes. Mann--Whitney U tests were used for univariate comparisons of continuous outcomes, and chi-square tests were used for binary outcomes. Regression analyses were used for the primary outcomes, with experimental condition as the main independent variable. Survey comparisons, learning-trajectory analyses, phenotype analyses, and process-correlation analyses were treated as exploratory.

\subsection{Data annotation}

In addition to the intervention study, we curated an open dataset of clinical history-taking dialogues from authentic instructional use of the platform. Because the control condition did not produce free-text dialogue, the corpus was drawn from the multi-agent learning and examination sessions, comprising 207 consultation sessions and 4{,}815 messages between medical students and the AI standardized patient (AI-SP). All transcripts were de-identified before annotation.

Each consultation was annotated under a four-dimension framework defined in a formal annotation manual, with each dimension applied at a different unit of analysis. The first dimension, AI-SP fidelity, was annotated on each patient message and recorded two binary judgments: script adherence, meaning whether the reply addressed the question and remained internally consistent, and progressive disclosure, meaning whether the patient disclosed information only when asked rather than prematurely. The second dimension, student dialogue intent, was annotated on each student message with two labels: a dialogue act from eight categories and a single information target from a thirty-item checklist whose history-taking items follow the OLDCARTS framework \cite{bickley2012bates}. The third dimension, need for scaffolding, was annotated on each student message as a binary judgment of whether tutor intervention was warranted, recording the main reason and recommended strategy when it was. The fourth dimension, global OSCE assessment, was rated once per session on the four domains described above.

Three clinical experts---two medical professors and one chief physician---independently annotated all 207 sessions, producing three parallel annotation tracks. The annotation manual provided shared label definitions, examples, unit-of-analysis rules, conservative thresholds for tutor intervention, and OSCE-domain anchors. The manual recommended a pilot calibration process on 10--15 sessions before formal annotation; formal annotation was then performed independently. Within each session, the experts read the full transcript, annotated the patient- and student-message dimensions turn by turn, and assigned the session-level OSCE ratings last to limit anchoring. The gold standard was the majority vote for nominal and binary labels and the mean of the three experts for OSCE ratings.

\subsection{Data analysis}

We summarized the corpus with descriptive statistics, reporting continuous variables as medians and ranges and categorical variables as frequencies and percentages, including the distributions of dialogue acts, information targets, scaffolding needs, and patient-fidelity outcomes.

We assessed inter-expert reliability separately for each dimension. For the nominal and binary constructs, we used Fleiss' \(\kappa\) \cite{fleiss1971measuring} across the three experts, with Krippendorff's \(\alpha\) \cite{krippendorff2018content} as a robustness check and the prevalence-adjusted and bias-adjusted \(\kappa\) (PABAK) \cite{byrt1993bias} for the highly imbalanced fidelity items. For the ordinal OSCE domains, we used the intraclass correlation coefficient (ICC) with absolute agreement, reporting ICC(2,k) for the three-expert mean used as the gold standard and ICC(2,1) for a single expert \cite{koo2016guideline}. We report label base rates alongside \(\kappa\) and describe agreement using the Landis--Koch bands \cite{landis1977measurement}. Analyses were performed in Python with NumPy \cite{harris2020array}, SciPy \cite{virtanen2020scipy}, and pingouin \cite{vallat2018pingouin}.

\section{Data Availability}

The multi-expert annotated dataset generated in this study will be made publicly available upon
acceptance of the manuscript and assigned a permanent DOI to ensure long-term accessibility. During
the peer-review process, the dataset can be made available to reviewers upon request by contacting
the corresponding author.

The dataset includes de-identified interview transcripts, structured checklist annotations, turn-level
evaluations, and OSCE-aligned scoring outcomes.

\section{Code Availability}
\label{sec:codeava}
The source code for the scaffolding-oriented multi-agent AI Standardized Patient (AI-SP) system is publicly available at: \url{https://github.com/skylynf/agent-medu}.
Upon acceptance of the manuscript, the repository will be archived and assigned a permanent DOI to ensure version control and reproducibility.
In addition, an online demonstration platform will be made openly accessible to allow researchers and educators to explore and interact with the system.



\clearpage

\bibliographystyle{unsrt}
\bibliography{meduai-sp}

\end{spacing}

\clearpage

\clearpage

\renewcommand{\thefigure}{S\arabic{figure}}
\renewcommand{\thetable}{S\arabic{table}}
\renewcommand{\theequation}{S\arabic{equation}}

\setcounter{figure}{0}
\setcounter{table}{0}
\setcounter{equation}{0}

\section*{Supplementary Information}

\setcounter{subsection}{0}
\renewcommand{\thesubsection}{S.\arabic{subsection}}

\begin{spacing}{1.1}


\subsection{Annotated Dialogue Examples}
\label{app:examples}

To illustrate the annotation scheme, we present representative excerpts in which
the three experts judged that tutor scaffolding was warranted after a student
utterance (the need-for-scaffolding dimension). Each box shows the local dialogue
context, the utterance that triggered the intervention judgment, and the expert
annotation, including the intervention reason, the recommended pedagogical
strategy, and an example of an ideal tutor prompt. Excerpts are translated from
the original Chinese; speaker turns are abbreviated for readability.

\newtcolorbox{dlgbox}[1]{%
  colback=gray!3, colframe=black!55, boxrule=0.5pt, arc=2pt,
  fonttitle=\bfseries\small, title={#1}, breakable,
  left=6pt, right=6pt, top=4pt, bottom=4pt}
\newcommand{\spk}[1]{\textbf{#1.}\,}
\newcommand{\flag}[1]{\colorbox{red!10}{\parbox{0.93\linewidth}{#1}}}

\begin{dlgbox}{Example 1 — Premature closure: a diagnosis is reached but the workup does not advance (acute pancreatitis); strategy: correction}
\small
\spk{Student} To confirm: you have persistent, knife-like upper-abdominal pain that worsens on pressure and when lying flat and eases slightly when leaning forward, radiating to the back in a band-like pattern and progressively worsening, with repeated vomiting, abdominal distension, no bowel movement, dark urine, and a low fever of 38.2\textdegree C. Is that correct?\\
\spk{Patient} Yes, doctor, exactly. It really hurts.\\
\spk{Student} Given the precipitants, the typical pain features, and the associated symptoms, I strongly suspect acute pancreatitis. I will complete the workup, confirm the diagnosis promptly, and treat it in time.\\
\spk{Patient} Okay doctor, please help me quickly; I am in so much pain.\\[2pt]
\flag{\spk{Student (intervention point)} Which is worse, daytime or nighttime?}
\tcblower
\small
\textbf{Need for scaffolding:} Yes \quad\textbf{Reason:} Premature closure \quad\textbf{Strategy:} Correction\\
\textbf{Ideal tutor prompt:} ``You have already summarized several high-risk features. Before asking lower-yield timing questions, what key tests or focused examination steps would help you confirm or rule out your leading diagnosis safely?''
\end{dlgbox}

\vspace{2em}

\begin{dlgbox}{Example 2 — Premature closure: leading questions before establishing the presentation (acute cholecystitis); strategy: hint}
\small
\emph{This was the student's opening utterance; the chief complaint, pain location, and onset had not yet been established.}\\[2pt]
\flag{\spk{Student (intervention point)} Does the pain radiate to the right shoulder or back? Is it worse after fatty food or a large meal?}
\tcblower
\small
\textbf{Need for scaffolding:} Yes \quad\textbf{Reason:} Premature closure \quad\textbf{Strategy:} Hint\\
\textbf{Ideal tutor prompt:} ``Before probing into specific features, first establish the basics, such as the chief complaint and the location and character of the pain. This lets you gather information systematically rather than testing a single hypothesis too early.''
\end{dlgbox}

\vspace{2em}
\begin{dlgbox}{Example 3 — Repetition / failure to use known information (acute cholecystitis); strategy: hint}
\small
\spk{Student} Does the pain radiate anywhere else, such as the right shoulder or back?\\
\spk{Patient} \emph{(answers, confirming radiation to the right shoulder and back)}\\[2pt]
\emph{[\,\ldots\ many turns later in the same consultation\,\ldots]}\\[2pt]
\spk{Student} What is your last menstrual period; are your cycles regular?\\
\spk{Patient} I am already postmenopausal at my age, so menstruation is not an issue.\\[2pt]
\flag{\spk{Student (intervention point)} Does the pain radiate to the right shoulder or back? \emph{(the same question already asked and answered earlier)}}
\tcblower
\small
\textbf{Need for scaffolding:} Yes \quad\textbf{Reason:} Stuck (redundant questioning) \quad\textbf{Strategy:} Hint\\
\textbf{Ideal tutor prompt:} ``You already asked about radiation to the right shoulder and back earlier in the encounter. Rather than repeating it, build on what you have already learned and move on to other associated symptoms or a focused examination.''
\end{dlgbox}

\vspace{2em}
\begin{dlgbox}{Example 4 — Critical error: distorting the history when summarizing (acute pancreatitis); strategy: correction}
\small
\emph{The patient had reported being woken by the pain at around 3--4 a.m., i.e., several hours earlier.}\\[2pt]
\spk{Student} What does the pain feel like? Is it intermittent, or constant without stopping?\\
\spk{Patient} It has been constant, never stopping, and getting worse.\\[2pt]
\flag{\spk{Student (intervention point)} Noted: persistent, knife-like pain in the mid-upper abdomen for half an hour, so severe you cannot straighten your back. Let me ask a few more details.}
\tcblower
\small
\textbf{Need for scaffolding:} Yes \quad\textbf{Reason:} Critical error (unconfirmed findings introduced) \quad\textbf{Strategy:} Correction\\
\textbf{Ideal tutor prompt:} ``The patient did not describe the pain as knife-like or say they could not straighten their back, and the pain has lasted since the early morning rather than for half an hour. Please restate the patient's information accurately and avoid adding unconfirmed symptoms.''
\end{dlgbox}

\vspace{2em}
\begin{dlgbox}{Example 5 — Critical error: requesting a result before the test is done (acute appendicitis); strategy: correction}
\small
\spk{Student} How long have you had the fever?\\
\spk{Patient} The fever started this morning; I measured 37.8\textdegree C.\\[2pt]
\flag{\spk{Student (intervention point)} Is the white-cell count abnormal?}
\tcblower
\small
\textbf{Need for scaffolding:} Yes \quad\textbf{Reason:} Critical error (data not yet available) \quad\textbf{Strategy:} Correction\\
\textbf{Ideal tutor prompt:} ``The patient has not had any blood test yet, so the white-cell count is not known. Should you first order a complete blood count?''
\end{dlgbox}

\vspace{2em}

\subsection{Case materials}
\label{supcasemat}

Three acute abdominal disease scenarios were constructed for the experiment: acute appendicitis (Figure \ref{supcase1acuteappend}), acute pancreatitis (Figure \ref{supcase2:acutepancrea}), and perforated peptic ulcer (Figure \ref{supcase3:perforatedpepulc}). Acute appendicitis and acute pancreatitis were used as learning cases, whereas perforated peptic ulcer was used as the examination case. All cases were encoded using a unified case structure that included patient profile, spontaneously volunteered information, inquiry-dependent responses, deeper symptom characterization, emotional behavior, pathophysiological background, expected diagnosis, and key differential diagnoses.

\begin{figure*}[htbp]
\caption{Supplementary case material for Acute appendicitis.}
\label{supcase1acuteappend}
\begin{tcolorbox}[
colframe=blue!40!gray,
colback=blue!5!gray!5,
colbacktitle=blue!40!gray,
fonttitle=\bfseries,
title=Supplementary Case Materials: Acute Abdominal Disease Scenarios,
rounded corners,
width=\textwidth,
boxrule=1pt
]

\vspace{1em}

\noindent \textbf{Case 1: Acute appendicitis.}\hspace{0.5em}\hrule height 0.3pt\hfill

\vspace{0.6em}
\textbf{Case role:} Learning case.

\vspace{0.4em}
\textbf{Patient profile.}
The patient was Zhang Ming, a 28-year-old man working as a software engineer. He was introverted and tended to answer only what was asked. On presentation, he appeared slightly pale and was bending forward while holding his abdomen.

\vspace{0.4em}
\textbf{Pathophysiological background.}
Acute appendicitis was modeled as obstruction of the appendiceal lumen, caused for example by a fecalith or lymphoid hyperplasia, leading to increased intraluminal pressure, impaired blood circulation in the appendiceal wall, bacterial invasion, and progressive inflammation. The inflammatory process may progress from simple appendicitis to suppurative appendicitis, gangrenous appendicitis, and perforation, potentially resulting in localized or diffuse peritonitis. A key clinical feature is migratory right lower-quadrant abdominal pain, typically beginning around the umbilicus and later localizing to McBurney's point.

\vspace{0.4em}
\textbf{Spontaneously volunteered information.}
At the beginning of the encounter, the patient stated: ``Doctor, my stomach hurts badly. I cannot stand it anymore.'' He also reported that the pain had started the previous night.

\vspace{0.4em}
\textbf{Inquiry-dependent responses.}
When asked about the location of pain, the patient reported that the pain initially started around the umbilicus and then gradually moved to the right lower abdomen. When asked about nausea or vomiting, he reported mild nausea and one episode of vomiting the previous night. When asked about fever, he stated that he felt slightly febrile and had measured a temperature of 37.8$^\circ$C that morning. When asked about bowel movements, he reported no bowel movement since the previous day. Urination was normal. Regarding the duration of symptoms, he stated that the pain began after dinner at approximately 8--9 p.m. the previous evening. When asked about potential triggers, he reported eating barbecue and drinking some beer with friends the previous night. He had no prior similar episodes, no history of surgery, no known drug allergies, and no chronic illnesses. His father had a history of gastric disease, but the rest of the family was healthy.

\vspace{0.4em}
\textbf{Deeper symptom characterization.}
When asked about the character of the pain, the patient described it as intermittent, dull pain at first, which later became continuous and progressively more severe. He reported that walking and coughing markedly worsened the right-sided abdominal pain, making him reluctant to move. The pain did not radiate elsewhere. There was no clear periodicity; it was persistent and more severe at night. He rated the pain as 7 out of 10. The onset was gradual rather than sudden, with progressive worsening.

\vspace{0.4em}
\textbf{Emotional and behavioral model.}
At baseline, the patient was anxious, frightened, and visibly distressed by pain. If the physician responded empathetically, he became more relaxed and was willing to provide more detailed answers. If the physician behaved coldly, he became more tense, gave shorter answers, and was less willing to elaborate. If the physician appeared rushed, he felt neglected and could omit deeper symptom information.

\vspace{0.4em}
\textbf{Expected diagnosis.}
Acute appendicitis, possibly suppurative appendicitis.

\vspace{0.4em}
\textbf{Key differential diagnoses.}
Acute gastroenteritis; right ureteral stone; Meckel's diverticulitis.

\vspace{1em}

\end{tcolorbox}
\end{figure*}

\begin{figure*}[htbp]
\caption{Supplementary case material for Acute pancreatitis.}
\label{supcase2:acutepancrea}
\begin{tcolorbox}[
colframe=blue!40!gray,
colback=blue!5!gray!5,
colbacktitle=blue!40!gray,
fonttitle=\bfseries,
title=Supplementary Case Materials: Acute Abdominal Disease Scenarios,
rounded corners,
width=\textwidth,
boxrule=1pt
]

\noindent \textbf{Case 2: Acute pancreatitis.}\hspace{0.5em}\hrule height 0.3pt\hfill

\vspace{0.6em}
\textbf{Case role:} Learning case.

\vspace{0.4em}
\textbf{Patient profile.}
The patient was Li Wei, a 45-year-old man working as a business owner. He was impatient and became irritable when in pain. On presentation, he appeared pale and diaphoretic and was bending forward while holding his knees.

\vspace{0.4em}
\textbf{Pathophysiological background.}
Acute pancreatitis was modeled as premature activation of pancreatic enzymes within the pancreas due to etiologies such as gallstones or heavy alcohol intake. This leads to autodigestion of pancreatic tissue, pancreatic edema, hemorrhage, and necrosis, with release of inflammatory mediators that may cause systemic inflammatory response syndrome and, in severe cases, multiple organ dysfunction syndrome. The typical presentation is persistent severe epigastric pain radiating to the lower back in a belt-like pattern, with partial relief in a forward-flexed or knee-chest position.

\vspace{0.4em}
\textbf{Spontaneously volunteered information.}
At the beginning of the encounter, the patient stated: ``Doctor, please help me quickly. My upper abdomen hurts terribly.'' He also reported that the pain was so severe that he could not straighten his back.

\vspace{0.4em}
\textbf{Inquiry-dependent responses.}
When asked about the location of pain, the patient pointed to the area above the umbilicus and stated that the entire upper abdomen hurt. He reported repeated vomiting, but vomiting did not relieve the pain. He had a fever of approximately 38.2$^\circ$C. He had not had a bowel movement that day and felt mildly bloated. His urine was yellow, but he was still able to urinate. Regarding the duration of symptoms, he stated that he was awakened by sudden pain at approximately 3--4 a.m. When asked about dietary or alcohol-related triggers, he reported drinking a substantial amount of liquor during a business dinner the previous night and eating many greasy dishes. He had previously been told during a health examination that he had gallstones, which were considered small and did not require treatment at that time. He had no history of surgery. He was unsure about drug allergies but believed he probably had none. He had hyperlipidemia but did not take medication regularly. His father had diabetes.

\vspace{0.4em}
\textbf{Deeper symptom characterization.}
When asked whether the pain radiated elsewhere, the patient reported that it radiated to his back and felt like a belt tightening around his waist. He described the pain as continuous, knife-like, and unbearable. He stated that bending forward and holding his knees provided slight relief, whereas lying flat made the pain worse. The pain was continuous without intermittent relief and had become progressively worse. He rated the pain as 9 out of 10. The onset was sudden, waking him from sleep in the middle of the night.

\vspace{0.4em}
\textbf{Emotional and behavioral model.}
At baseline, the patient was restless, extremely distressed, and fearful. If the physician responded empathetically, he felt somewhat reassured, although the pain persisted and his speech remained urgent. If the physician behaved coldly, he became more irritable and repeatedly urged the physician to relieve his pain. If the physician appeared rushed, his anxiety worsened and his responses became disorganized.

\vspace{0.4em}
\textbf{Expected diagnosis.}
Acute pancreatitis, possibly biliary in origin and triggered by alcohol intake.

\vspace{0.4em}
\textbf{Key differential diagnoses.}
Perforated peptic ulcer; acute cholecystitis; acute myocardial infarction, especially inferior-wall myocardial infarction.

\vspace{1em}

\end{tcolorbox}
\end{figure*}

\begin{figure*}[htbp]
\caption{Supplementary case material for Perforated peptic ulcer.}
\label{supcase3:perforatedpepulc}
\begin{tcolorbox}[
colframe=blue!40!gray,
colback=blue!5!gray!5,
colbacktitle=blue!40!gray,
fonttitle=\bfseries,
title=Supplementary Case Materials: Acute Abdominal Disease Scenarios,
rounded corners,
width=\textwidth,
boxrule=1pt
]

\noindent \textbf{Case 3: Perforated peptic ulcer.}\hspace{0.5em}\hrule height 0.3pt\hfill

\vspace{0.6em}
\textbf{Case role:} Examination case.

\vspace{0.4em}
\textbf{Patient profile.}
The patient was Wang Jianguo, a 52-year-old man working as a taxi driver. He had high pain tolerance and was accustomed to enduring discomfort. On presentation, he appeared ashen, diaphoretic, and lay supine, avoiding movement.

\vspace{0.4em}
\textbf{Pathophysiological background.}
Perforated peptic ulcer was modeled as chronic injury to the gastric or duodenal mucosa caused by gastric acid and pepsin, resulting in ulcer formation. When the ulcer erodes through the serosal layer, perforation occurs, allowing gastrointestinal contents to enter the abdominal cavity. This initially causes chemical peritonitis and may subsequently lead to bacterial peritonitis. The typical presentation is sudden, severe, knife-like epigastric pain that rapidly spreads to the entire abdomen, accompanied by board-like abdominal rigidity and evidence of pneumoperitoneum.

\vspace{0.4em}
\textbf{Spontaneously volunteered information.}
At the beginning of the encounter, the patient stated: ``Doctor, my upper abdomen suddenly became terribly painful, as if I had been stabbed with a knife.'' He also reported that he did not dare to move.

\vspace{0.4em}
\textbf{Inquiry-dependent responses.}
When asked about the location of pain, the patient stated that it began in the upper abdomen and had now spread to the entire abdomen. He felt nauseated but had not vomited. He reported feeling somewhat chilled, but his temperature had not yet been measured. He had not had a bowel movement that day. Urination was normal. Regarding the duration of symptoms, he stated that the pain had started suddenly one hour earlier, shortly after lunch. When asked about food intake, he reported eating only a few hurried bites of lunch and drinking a glass of cold water. He had a history of chronic gastric disease for more than ten years, with recurrent upper abdominal pain that usually improved after taking medication. He had no history of surgery and was unsure about drug allergies, but believed he probably had none. He had hypertension and was taking antihypertensive medication. He reported no notable family history. He had smoked for more than 20 years, approximately one pack per day, and drank alcohol occasionally.

\vspace{0.4em}
\textbf{Deeper symptom characterization.}
When asked about the character of the pain, the patient described it as if a knife were stabbing inside his abdomen and stated that he had never experienced such severe pain before. The onset was extremely sudden, with intense pain beginning almost instantaneously. Movement, deep breathing, and turning over worsened the pain, so he remained still and did not dare to move. He also reported vague discomfort in the right shoulder. He rated the pain as 10 out of 10. The pain was continuous and had not stopped. He reported that he usually took omeprazole for gastric pain, but had not had time to take it on this occasion.

\vspace{0.4em}
\textbf{Emotional and behavioral model.}
At baseline, the patient was in extreme pain but restrained and did not shout loudly. If the physician responded empathetically, he felt grateful and tried to cooperate with the interview. If the physician behaved coldly, he became silent and gave only brief answers. If the physician appeared rushed, he became nervous but still tried to cooperate despite the pain.

\vspace{0.4em}
\textbf{Expected diagnosis.}
Perforated peptic ulcer, likely perforation of a duodenal bulb ulcer.

\vspace{0.4em}
\textbf{Key differential diagnoses.}
Acute pancreatitis; perforated acute cholecystitis; acute myocardial infarction.

\vspace{2mm}

\end{tcolorbox}
\end{figure*}

\begin{table}[!t]
\centering
\footnotesize
\setlength{\tabcolsep}{4pt}
\renewcommand{\arraystretch}{1.02}

\caption{\textbf{Participant characteristics and experimental workflow.}}
\label{tab:participants}

\begin{threeparttable}
\begin{tabularx}{\linewidth}{
>{\raggedright\arraybackslash}p{0.25\linewidth}
>{\raggedright\arraybackslash}p{0.33\linewidth}
>{\raggedright\arraybackslash}p{0.17\linewidth}
>{\raggedright\arraybackslash}X
}
\toprule
\textbf{Section} & \textbf{Item} & \textbf{Statistic or unit} & \textbf{Value} \\
\midrule

\multicolumn{4}{l}{\textbf{Participant demographics}} \\

Institution
& Guangzhou Medical University
& --
& Undergraduate clinical-medicine program \\

Total enrolled
& Volunteer participants
& $n$
& 100 \\

Age
& Range; mean
& Years
& 20--26; 22.3 \\

Sex
& Women; men
& $n$
& 63; 37 \\

Biological materials
& Human specimens collected
& --
& None \\

Patient data
& Real clinical patient data collected
& --
& None \\

\midrule
\multicolumn{4}{l}{\textbf{Study completion and analytic sample}} \\

Completed full workflow
& Participants finishing all stages
& $n$
& 95 \\

Primary analysis
& Complete-case analysis
& $n$
& 95 \\

Multi-agent group (MA)
& Complete cases
& $n$
& 47 \\

Control group (CT)
& Complete cases
& $n$
& 48 \\

Attrition
& Incomplete workflow
& $n$
& 5 \\

\bottomrule
\end{tabularx}

\begin{tablenotes}[flushleft]
\footnotesize
\item Note: Participants were volunteer learners. The primary analysis followed a complete-case approach including only participants who completed the full experimental workflow. No biological samples or identifiable patient clinical data were collected.
\end{tablenotes}
\end{threeparttable}
\end{table}

\begin{table}[!t]
\centering
\footnotesize
\setlength{\tabcolsep}{4pt}
\renewcommand{\arraystretch}{1.02}

\caption{\textbf{MeduAI-SP platform architecture and statistical analysis environment.}}
\label{tab:platform_environment}

\begin{threeparttable}
\begin{tabularx}{\linewidth}{l l c X}
\toprule
\textbf{Section} & \textbf{Component} & \textbf{Version / tool} & \textbf{Description} \\
\midrule

\multicolumn{4}{l}{\textbf{Web-based MeduAI-SP platform}} \\

Frontend framework
& React
& 18.3.1
& User interface for case selection, dialogue, examination, and questionnaires \\

Build tool
& Vite
& 6.0.5
& Frontend bundling and development server \\

Programming language (FE)
& TypeScript
& 5.7.2
& Static typing and development support \\

Styling
& TailwindCSS
& 3.4.17
& Utility-first CSS framework \\

Visualization
& Recharts
& 2.15.0
& Interactive charts and analytics dashboard \\

Backend language
& Python
& 3.11
& Core server-side implementation \\

Backend framework
& FastAPI
& 0.115.6
& RESTful API and WebSocket services \\

ORM
& SQLAlchemy
& 2.0.36
& Database abstraction layer \\

Async driver
& asyncpg
& 0.30.0
& PostgreSQL asynchronous driver \\

WebSocket library
& websockets
& 14.1
& Real-time dialogue communication \\

YAML parser
& PyYAML
& 6.0.2
& Prompt template and case configuration loading \\

Database
& PostgreSQL
& 15
& Storage of sessions, messages, evaluations, and surveys \\

Large language model
& qwen-max 
& --
& AI-SP, tutor, and evaluator agents \\

\midrule
\multicolumn{4}{l}{\textbf{Statistical analysis environment}} \\

Core numerical library
& NumPy
& 1.26.4
& Numerical computation \\

Data processing
& pandas
& 2.2.2
& Data cleaning and tabulation \\

Statistical tests
& SciPy
& 1.13.1
& Classical statistical inference \\

Regression modeling
& statsmodels
& 0.14.2
& Generalized linear models and inference \\

Machine learning
& scikit-learn
& 1.5.1
& Classification, clustering, and metrics \\

Visualization
& Matplotlib
& 3.9.2
& Statistical plotting \\

Reliability analysis
& Pingouin
& --
& Intraclass correlation coefficients (ICC) \\

Dimensionality reduction
& umap-learn
& --
& UMAP embedding \\

Clustering
& scikit-learn (k-means)
& 1.5.1
& Unsupervised clustering \\

\bottomrule
\end{tabularx}

\begin{tablenotes}[flushleft]
\footnotesize
\item Note: Dialogue records, timestamps, worksheets, agent outputs, evaluation results, questionnaire responses, and prompt-version snapshots were automatically recorded by the MeduAI-SP platform.
\end{tablenotes}
\end{threeparttable}
\end{table}

\begin{table}[!t]
\centering
\footnotesize
\setlength{\tabcolsep}{5pt}
\renewcommand{\arraystretch}{1.08}

\caption{\textbf{Examination outcomes: descriptive statistics and univariate group comparisons.}}
\label{tab:s3_exam_outcomes}

\begin{threeparttable}
\begin{tabularx}{\linewidth}{
>{\raggedright\arraybackslash}X
>{\raggedright\arraybackslash}p{0.22\linewidth}
>{\centering\arraybackslash}p{0.16\linewidth}
>{\centering\arraybackslash}p{0.18\linewidth}
}
\toprule
\textbf{Variable / Outcome} & \textbf{Statistical test} & \textbf{\(P\) value} & \textbf{Hedges' \(g\) (approx)} \\
\midrule

\multicolumn{4}{l}{\textbf{Continuous outcomes}} \\

final\_score & Mann--Whitney U & \(5.51 \times 10^{-5}\) & -0.814 \\
osce\_communication & Mann--Whitney U & \(4.14 \times 10^{-4}\) & -0.790 \\
weighted\_score\_pct & Mann--Whitney U & \(0.039\) & -0.311 \\
completion\_rate & Mann--Whitney U & \(0.045\) & -0.326 \\
items\_checked & Mann--Whitney U & \(0.045\) & -0.326 \\
osce\_total / osce\_mean & Mann--Whitney U & \(0.095\) & -0.331 \\
osce\_history\_completeness & Mann--Whitney U & \(0.106\) & -0.354 \\
osce\_clinical\_reasoning & Mann--Whitney U & \(0.893\) & 0.004 \\
osce\_diagnostic\_accuracy & Mann--Whitney U & \(0.717\) & 0.038 \\
duration\_seconds & Mann--Whitney U & \(0.599\) & -0.040 \\
missed\_critical & Mann--Whitney U & \(0.773\) & 0.052 \\
n\_differentials\_given & Mann--Whitney U & \(0.512\) & -0.108 \\

\midrule
\multicolumn{4}{l}{\textbf{Binary outcomes}} \\

diagnosis\_correct & $\chi^2$ & \(1.000\) & -- \\
worksheet\_filled & $\chi^2$ & \(0.729\) & -- \\

\bottomrule
\end{tabularx}

\begin{tablenotes}[flushleft]
\footnotesize
\item Note: Continuous outcomes were compared using Mann--Whitney U tests. Binary outcomes were analyzed using chi-square tests. Negative Hedges' \(g\) values indicate higher scores in the MA group relative to the CT group (export convention).
\end{tablenotes}

\end{threeparttable}
\end{table}

\begin{table}[!t]
\centering
\footnotesize
\setlength{\tabcolsep}{6pt}
\renewcommand{\arraystretch}{1.1}

\caption{\textbf{Internal consistency reliability of study questionnaires (Cronbach's $\alpha$).}}
\label{tab:s1_reliability}

\begin{threeparttable}
\begin{tabularx}{\linewidth}{
>{\raggedright\arraybackslash}l
>{\raggedright\arraybackslash}X
>{\centering\arraybackslash}p{0.12\linewidth}
>{\centering\arraybackslash}p{0.12\linewidth}
>{\centering\arraybackslash}p{0.14\linewidth}
>{\raggedright\arraybackslash}X
}
\toprule
\textbf{Instrument} & \textbf{Scope} & $\boldsymbol{\alpha}$ & \textbf{Items} & \textbf{Respondents} & \textbf{Interpretation} \\
\midrule

SUS & Overall & 0.856 & 10 & 70 & Good \\

\midrule

UES & Overall & 0.958 & 30 & 53 & Excellent \\
UES & Focused Attention (FA) & 0.946 & 7 & 60 & Excellent \\
UES & Perceived Usability (PU) & 0.867 & 8 & 59 & Good \\
UES & Aesthetic Appeal (AE) & 0.943 & 5 & 61 & Excellent \\
UES & Reward Factor (RW) & 0.947 & 10 & 60 & Excellent \\

\bottomrule
\end{tabularx}

\begin{tablenotes}[flushleft]
\footnotesize
\item Note: Cronbach's $\alpha$ values indicate internal consistency reliability. Interpretation follows conventional thresholds (0.70–0.79 acceptable; 0.80–0.89 good; $\geq$0.90 excellent).
\end{tablenotes}

\end{threeparttable}
\end{table}

\subsection{Supplementary assessment and intervention details}

\begin{table}[!t]
\centering
\footnotesize
\setlength{\tabcolsep}{4pt}
\renewcommand{\arraystretch}{1.15}
\caption{\textbf{OSCE-aligned global scoring rubric.}}
\label{tab:s5_osce_rubric}
\begin{threeparttable}
\begin{tabularx}{\linewidth}{>{\raggedright\arraybackslash}p{0.20\linewidth}>{\raggedright\arraybackslash}p{0.23\linewidth}>{\raggedright\arraybackslash}p{0.25\linewidth}>{\raggedright\arraybackslash}X}
\toprule
\textbf{Domain} & \textbf{Score 1 anchor} & \textbf{Score 3 anchor} & \textbf{Score 5 anchor} \\
\midrule
History taking & Clearly insufficient; disorganized questioning with major omissions in chief complaint, present illness, or relevant background history & Meets basic expectations; covers core symptoms and some associated or background items but with gaps or inefficient sequencing & Excellent; systematic, prioritized, and complete coverage of chief complaint, OLDCARTS features, associated symptoms, red flags, and relevant past/family/social history \\
Clinical thinking & Little evidence of hypothesis generation; major premature closure or unfocused questioning & Basic clinical hypothesis is present; some verification or exclusion questions are asked, but reasoning remains incomplete & Clear hypothesis-driven reasoning; uses targeted questions, examination, and tests to compare diagnoses and avoid premature closure \\
Communication and empathy & Cold, unclear, or disruptive communication; little response to patient distress or confusion & Professional and understandable communication with occasional empathy or confirmation but inconsistent rapport building & Patient-centered communication; clear explanations, self-introduction, reassurance, empathy, respect for patient concerns, and effective summarization/confirmation \\
Diagnostic accuracy & Final diagnosis absent, unsafe, or inconsistent with the case information & Plausible primary diagnosis or differential but incomplete justification or management alignment & Correct primary diagnosis with appropriate differentials and a management or workup plan consistent with the diagnostic direction \\
\bottomrule
\end{tabularx}
\begin{tablenotes}[flushleft]
\footnotesize
\item Note: All domains used the shared 1--5 anchors: 1 = clearly insufficient, 2 = poor, 3 = meets basic expectations, 4 = good, and 5 = excellent. Expressing empathy contributed to both the item-level checklist and the communication-and-empathy global domain.
\end{tablenotes}
\end{threeparttable}
\end{table}

\begin{table}[!t]
\centering
\footnotesize
\setlength{\tabcolsep}{4pt}
\renewcommand{\arraystretch}{1.15}
\caption{\textbf{Examples of tutor scaffolding prompts used to guide clinical interviewing.}}
\label{tab:s6_prompt_examples}
\begin{threeparttable}
\begin{tabularx}{\linewidth}{>{\raggedright\arraybackslash}p{0.20\linewidth}>{\raggedright\arraybackslash}p{0.26\linewidth}>{\raggedright\arraybackslash}X}
\toprule
\textbf{Trigger} & \textbf{Scaffolding goal} & \textbf{Example prompt} \\
\midrule
Missing key history & Focused history taking & ``You have asked where the pain is. What other pain characteristics would help you judge severity and possible causes, such as onset, course, radiation, or aggravating factors?'' \\
Premature closure & Clinical reasoning & ``Is the information you have collected sufficient to support your leading diagnosis? What important alternatives or red flags should you still check before deciding?'' \\
Communication breakdown & Empathic communication & ``The patient has just expressed severe pain and worry. Before continuing with more questions, how could you briefly acknowledge the discomfort and reassure the patient that you are listening?'' \\
Redundant questioning & Summarization and confirmation & ``You have already obtained several key details. Consider summarizing what you know so far and then moving to the next missing area instead of repeating the same question.'' \\
\bottomrule
\end{tabularx}
\begin{tablenotes}[flushleft]
\footnotesize
\item Note: Prompts were phrased as Socratic questions or brief hints. The tutor was instructed not to directly disclose the final diagnosis or provide the case answer.
\end{tablenotes}
\end{threeparttable}
\end{table}

\begin{table}[!t]
\centering
\footnotesize
\setlength{\tabcolsep}{3pt}
\renewcommand{\arraystretch}{1.12}
\caption{\textbf{Alignment between case scenarios and scoring constructs.}}
\label{tab:s7_case_alignment}
\begin{threeparttable}
\begin{tabularx}{\linewidth}{>{\raggedright\arraybackslash}p{0.18\linewidth}>{\raggedright\arraybackslash}p{0.29\linewidth}>{\raggedright\arraybackslash}p{0.22\linewidth}>{\raggedright\arraybackslash}X}
\toprule
\textbf{Case} & \textbf{Key history and communication items} & \textbf{Expected diagnosis} & \textbf{Important differential diagnoses} \\
\midrule
Acute appendicitis & Migratory abdominal pain, right lower-quadrant location, onset and duration, nausea or vomiting, fever, bowel and urinary symptoms, pain aggravation with movement, prior episodes, allergy and medication history, empathy for pain and anxiety & Acute appendicitis & Gastroenteritis; right ureteral stone; ectopic pregnancy; Meckel's diverticulitis \\
Acute pancreatitis & Severe persistent epigastric pain, belt-like radiation to the back, alcohol or fatty-meal trigger, vomiting without relief, fever, abdominal distension or bowel changes, gallstone history, hyperlipidemia, posture-related relief, medication or allergy history, empathic response to severe distress & Acute pancreatitis, possibly biliary in origin and triggered by alcohol intake & Perforated peptic ulcer; acute cholecystitis; acute myocardial infarction, especially inferior-wall myocardial infarction \\
Perforated peptic ulcer & Sudden severe upper-abdominal pain, board-like abdominal rigidity, prior ulcer symptoms, NSAID or medication history, vomiting or hematemesis, fever or systemic symptoms, bowel changes, peritoneal signs, risk factors, prompt acknowledgement of fear and pain & Perforated peptic ulcer & Acute pancreatitis; acute appendicitis; acute cholecystitis; intestinal obstruction; ruptured abdominal aortic aneurysm \\
\bottomrule
\end{tabularx}
\begin{tablenotes}[flushleft]
\footnotesize
\item Note: Case-specific items were mapped to the same OSCE-aligned constructs: systematic history taking, clinical thinking, communication and empathy, diagnostic accuracy, and weighted checklist coverage.
\end{tablenotes}
\end{threeparttable}
\end{table}

\end{spacing}

\end{document}